\documentclass[useAMS,usenatbib]{mn2e}

\def\Zsol{\hbox{Z$_{\odot}$}}
\def\Msol{\hbox{M$_{\odot}$}}

\usepackage{epsfig,natbib,url,graphicx,subfig}
\usepackage{lscape,amsmath,amssymb}
\usepackage{booktabs,threeparttable}
\usepackage{longtable,deluxetable}
\usepackage{color}
\usepackage{soul} 
\usepackage{fancyvrb}
\usepackage{pifont}
\usepackage{ulem}
\usepackage{pdflscape}

\newcommand{\hi}{H\,{\sc i}}
\newcommand{\hii}{H~{\sc ii}}
\newcommand{\hei}{He~{\sc i}}
\newcommand{\heii}{He~{\sc ii}}
\newcommand{\kms}{km\,s$^{-1}$}
\newcommand{\eld}{$N_{\rm e}$}

\newcommand{\elt}{$T_{\rm e}$}

\newcommand{\Sp}{S$^+$}

\newcommand{\op}{O$^+$}

\newcommand{\opp}{O$^{2+}$}

\newcommand{\neppp}{Ne$^{3+}$}

\newcommand{\arppp}{Ar$^{3+}$}
\newcommand{\foiii}{[O~{\sc iii}]}

\newcommand{\foi}{[O~{\sc i}]}
\newcommand{\foii}{[O~{\sc ii}]}
\newcommand{\fsii}{[S~{\sc ii}]}
\newcommand{\fsiii}{[S~{\sc iii}]}

\newcommand{\fnii}{[N~{\sc ii}]}

\newcommand{\fneiii}{[Ne~{\sc iii}]}

\newcommand{\fariii}{[Ar~{\sc iii}]}
\newcommand{\ariv}{Ar~{\sc iv}}

\newcommand{\feiii}{Fe~{\sc iii}}

\newcommand{\hp}{H$^+$}

\newcommand{\ha}{H$\alpha$}
\newcommand{\hb}{H$\beta$}
\newcommand{\hg}{H$\gamma$}
\newcommand{\hd}{H$\delta$}

\title[Blue Diffuse Dwarf Galaxies]{Uncovering Blue Diffuse Dwarf Galaxies}
\author[James et al. ]{Bethan L. James$^{1}$\thanks{E-mail:bjames@ast.cam.ac.uk}, Sergey Koposov$^{1}$, Daniel P. Stark$^{2}$, Vasily Belokurov$^{1}$,  
\newauthor Max Pettini$^{1}$ \& Edward W. Olszewski$^2$\\
$^{1}$Institute of Astronomy, University of Cambridge, Madingley Road, Cambridge, CB3 0HA\\
$^{2}$Steward Observatory, The University of Arizona, 933 N Cherry Ave, Tucson, AZ, 85721, USA}
\begin{document}

\date{Accepted 2015 January 23. Received  in original form 2014 October 3}

\pagerange{\pageref{firstpage}--\pageref{lastpage}} \pubyear{2014}

\maketitle

\label{firstpage}

\begin{abstract}
Extremely metal poor (XMP) galaxies are known to be very rare, despite the large numbers of low-mass galaxies predicted by the local galaxy luminosity function. This paper presents a sub-sample of galaxies that were selected via a morphology-based search on SDSS images with the aim of finding these elusive XMP galaxies.  By using the recently discovered extremely metal-poor galaxy, Leo~P, as a guide, we obtained a collection of faint, blue systems, each with isolated \hii\ regions embedded in a diffuse continuum, that have remained optically undetected until now.  Here we show the first results from optical spectroscopic follow-up observations of 12 of $\sim$100 of these blue, diffuse dwarf (BDD) galaxies yielded by our search algorithm.   Oxygen abundances were obtained via the direct method for eight galaxies, and found to be in the range $7.45<12+\log{\rm (O/H)} <8.0$, with two galaxies being classified as XMPs.  All BDDs were found to currently have a young star-forming population ($<10$~Myr) and relatively high ionisation parameters
of their \hii\ regions. Despite their low luminosities ($-11\lesssim M_B \lesssim -18$) and low surface brightnesses ($\sim 23$--25\,mag~arcsec$^{-2}$), the galaxies were found to be actively star-forming, with current star-formation rates between 0.0003 and 0.078~\Msol\,yr$^{-1}$. From our current subsample, BDD galaxies appear to be a population of non-quiescent dwarf irregular (dIrr) galaxies, or the diffuse counterparts to blue compact galaxies (BCDs) and as such may bridge the gap between these two populations.  Our search algorithm demonstrates that morphology-based searches are successful in uncovering more \textit{diffuse} metal-poor star-forming galaxies, which traditional emission-line based searches overlook.
\end{abstract}


\section{Introduction}

In order to further our understanding of the early Universe, we continually search for galaxies whose gas shows little or no trace of chemical evolution.  Exploration of such pristine environments can provide insight into early galaxy formation and the properties of low-metallicity stars, along with constraints on theories of stellar evolution, the early enrichment of the pre-galactic medium, and studies of primordial elements \citep{Skillman:2013}.  Advancements continue to be made in this field, as searches now extend back to 400~Myr after the Big Bang \citep{Coe:2013} and candidates with photometric redshifts between $z\sim9-10$ are frequently being detected \citep[e.g.][]{Zheng:2012,Ellis:2013,Oesch:2013}.  However, due to their intrinsic faintness, very limited constraints can be extracted from such systems and we instead rely on the most unenriched galaxies in the \textit{local} Universe to provide a valuable means of constraining the early star formation, and consequently complement the ongoing high-$z$ searches.

Unfortunately metal-poor objects are very rare, especially those belonging to the class of \textit{extremely} metal poor
(XMP) galaxies (defined as having 12+log(O/H)$<$7.65, less than one tenth solar using the solar oxygen abundance of \citet{Asplund:2009}). 
For many decades, I~Zw18 \citep[12+log(O/H)=7.18$\pm$0.03 or 1/31~\Zsol,][]{Legrand:2000} and SBS~0335-052W \citep[12+log(O/H)=7.12$\pm$0.03 or 1/37~\Zsol,][]{Izotov:2005} have held the
record for the lowest metallicities and, at present, only 129 XMPs can
be found in the literature \citep[see][for a compilation]{Morales-Luis:2011}. Whilst their observed rarity is in part due to their inherent low-luminosities \citep[as demonstrated
by the luminosity-metallicity (L--Z) relation e.g.][]{Lequeux:1979,Skillman:1989,Garnett:2002}, it is also in contradiction with the large number of low-mass galaxies predicted by the local galaxy luminosity function.   This suggests that a large population of XMP galaxies still remains undetected.

The search for XMP galaxies has thus far focused primarily on surveying emission-line galaxies. An up-to-date summary of such surveys can be found in \citet{Skillman:2013}, from which it is clear that although considerable effort has been spent in increasing the number of XMPs, the resultant yield has been very low. For example, of one million 
Sloan Digital Sky Survey (SDSS) DR-7 spectra, there are only 15 candidates with 12+log(O/H)$<$7.35 within the Local Volume (Izotov et al. 2012). The low yield from these spectroscopic surveys is partly due to their design: they require naturally low-surface brightness objects to have high surface brightness \hii\ regions for detectability and abundance analysis.

An alternative strategy is to instead base the search criteria on L--Z relationship itself.  Surveys of nearby, low-luminosity galaxies, and blind \hi\ surveys have both been successful in detecting several XMPs \citep[e.g.][]{Skillman:1989,VanZee:2000,Pustilnik:2005,Cannon:2011,Haynes:2011,Tollerud:2015}, and in fact it was one discovery of this kind that gave rise to the sample of galaxies presented here.  Via its \hi\ signature alone, Leo~P, a nearby dwarf irregular galaxy, was recently brought to light through the ALFALFA survey \citep{Giovanelli:2013}.  Whilst its \hi\ properties were consistent with it being a `ultra-compact high velocity cloud', optical spectroscopic follow-up by \citet{Skillman:2013} revealed that Leo~P was in fact one of the most metal-poor galaxies ever observed, with 12+log(O/H)$=7.17\pm0.04$ ($\sim$1/34~\Zsol).  

Along with its extremely low metallicity, Leo~P's diffuse morphology and low-surface brightness \citep[$\mu_v=24.5\pm0.6$~mag arcsec$^{-2}$,][]{McQuinn:2013} are atypical of most XMPs.  The majority of low-metallicity galaxies are compact in nature, with only one or two bright \hii\ regions dominating their structure and are often found to have a cometary or tadpole-shaped morphology \citep{Papaderos:2008,Morales-Luis:2011,Sanchez-Almeida:2013}.  Leo~P, on the other hand, consists of randomly distributed \hii\ regions embedded within very low surface brightness emission.  This lead us to speculate about how common this unconventional morphology may be amongst XMPs and whether we should be focussing on morphology rather than emission line properties when searching optical surveys for XMPs. 

\begin{figure*}
\includegraphics[scale=0.5]{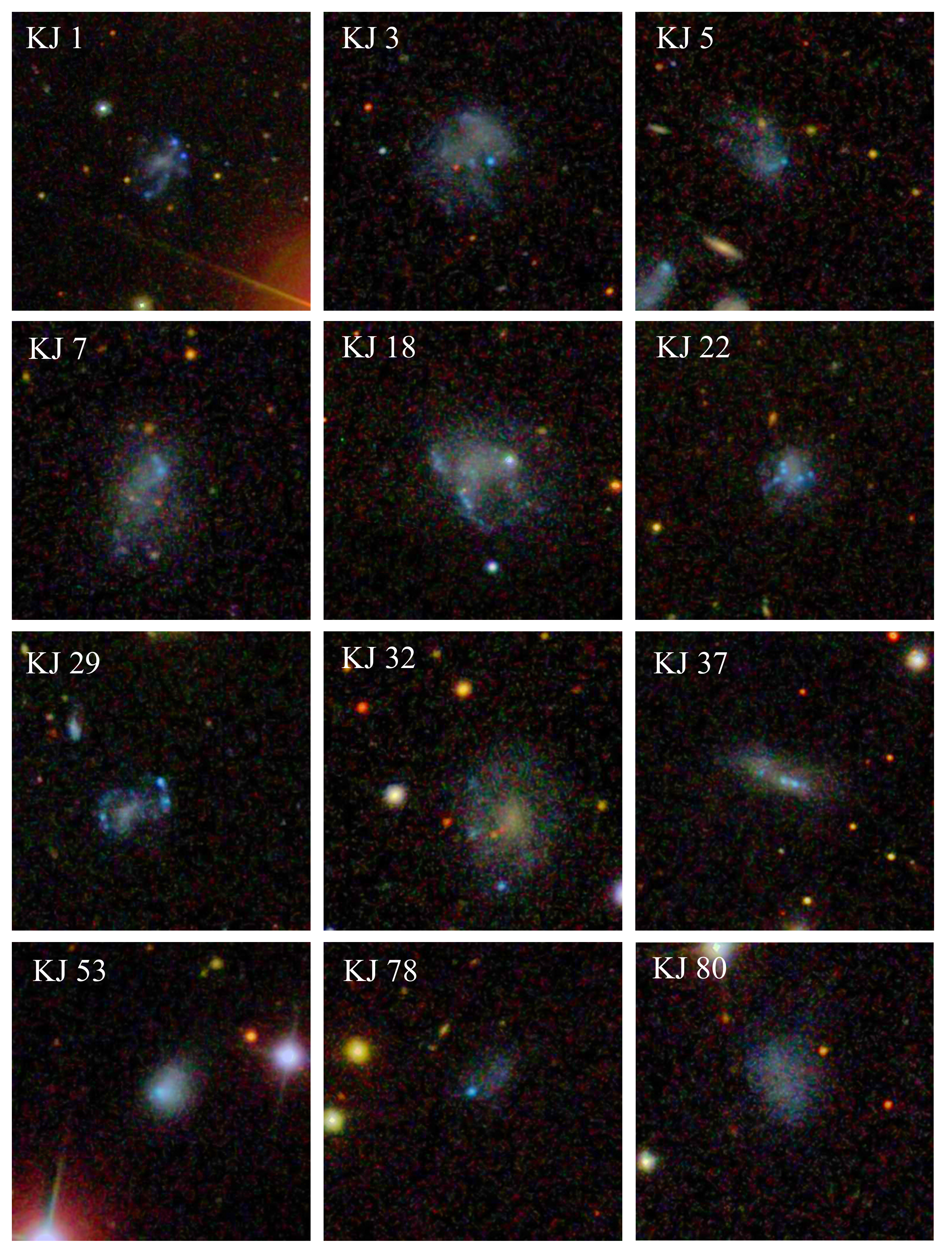}
\caption{SDSS images of the blue diffuse dwarf galaxies presented in this paper.  Information on each object can be found in Table~\ref{tab:gals}. Images were created from $gri$-band SDSS imaging and are 
$\sim 102$\,\arcsec on each side. 
North is up and east is to the left. }
\label{fig:images}
\end{figure*}

With the aim of answering this question, we undertook a systematic search in the Sloan source catalogue based on the morphology of Leo~P (full details are given below).  The result is a collection of faint, blue systems, each with isolated \hii\ regions embedded in a diffuse continuum.  We show a sub-sample of these `blue diffuse dwarfs' (BDDs) in Figure~\ref{fig:images}. In this publication, we present the first results from our ongoing optical spectroscopic follow-up of the sample.  Whilst our spectroscopic survey is far from complete, we take this opportunity to introduce this previously optically undetected sample of galaxies to the community via their emission line and morphological properties.
\begin{table*}
\caption{Blue diffuse dwarfs galaxies in the current sample.  Identifications according to SDSS and AGC are also given where available.   Redshifts were derived from  fitting all the emission lines seen within the spectrum and converted into distances using $H_0=67.8$~\kms\,Mpc$^{-1}$ \citep{Planck:2014}.
SDSS images of each target can be found in Figure 1.  PA and \textit{T} refer to the position angle of the slit orientation and the exposure time of the MMT observations, respectively.}
\begin{center}
\begin{tabular}{ccccccccc}
\hline
  Obj ID & SDSS ID& AGC ID & R.A. & Dec & $z$ & D (Mpc) & PA ($\deg$) & $T_{\rm exp}$ (s) \\
 \hline
KJ~1 & J160753.65+090703.7 & 268137 & 16:07:53.652 &  09:07:03.76 & 0.02796 & 123 & 135 & 2$\times$ 600\\ 
KJ~3 &  J141708.66+134105.5 &  242011 & 14:17:08.285 &13:41:08.54   & 0.01611 & 72 & 20 & 2$\times$ 1200\\
KJ~5 & J085739.15+590256.4 & --- &  08:57:39.155 &59:02:56.41   & 0.00374 & 16 & 315 & 2$\times$ 1200\\ 
KJ~7 & J113453.11+110116.1  &  212838  &11:34:53.455 &11:01:11.16   & 0.00249 & 11 & 223 &2$\times$ 1200\\ 
KJ~18 & J092127.17+072152.7 &   193816 & 09:21:27.173 &07:21:50.70   & 0.00414 & 18 & 24 &2$\times$ 1200\\
KJ~22  &  J114109.10+244505.2&  749439 & 11:41:08.885 &24:45:04.75   & 0.01146 & 51 &114 & 2$\times$ 1200\\
KJ~29  & J130110.39+365424.2 & --- & 13:01:11.229 &36:54:14.74   & 0.02252 & 100 &105& 1$\times$ 950\\
KJ~32 & J021111.29+281618.9 & 123128 & 02:11:10.356 & 28:16:09.84 & 0.01436 & 63 & 175 & 2$\times$ 1200\\

KJ~37  &  J062538.13+655228.0 & ---& 06:25:38.140 &65:52:28.31   & 0.01387 & 61 &20 & 3$\times$ 1200\\
KJ~53  & J115754.21+563816.7 & ---& 11:57:54.213 &56:38:16.76   & 0.00107 & 5& 135 &2$\times$ 1200\\
 KJ~78  & J092036.41+494031.3 & ---& 09:20:36.480 &49:40:30.94   & 0.00159 & 7&  64 & 3$\times$ 1200\\
 KJ~80  & J092603.64+560915.5 & ---& 09:26:03.600 &56:09:15.48   & 0.00251 & 11&73 & 1$\times$ 1200\\
  \hline
\end{tabular}
\end{center}
\label{tab:gals}
\end{table*}%

The paper is structured as follows. In Section~2 we describe our methodology for generating the catalogue of BDDs.  In Section~3 we provide information concerning the optical spectroscopic follow-up data, which are then presented in Section~4.  In Section 5 we use the spectra to calculate the chemical abundances for each target, along with several emission-line properties in Section~6 (i.e. star-formation rate, luminosity, ionisation parameter, and age of the current stellar population). In Section~7 we use these properties to assess where BDDs lie in relation to current dwarf populations and discuss the morphological properties of the sample in Section 8.  In Section 9 we summarise our findings and discuss the nature of BDDs and present our conclusions in Section~10. 

\section{Selection of galaxies}

The selection strategy of galaxies for this paper was primarily motivated by
the morphology of recently discovered Leo~P dwarf galaxy.  While discovered
using radio data, the galaxy is actually visible in the SDSS images as a
very diffuse low surface brightness cloud with a couple of stellar-like HII
regions.  Thus we have devised a search method which looks for
objects which are morphologically similar to Leo~P in the photometric SDSS database. For 
this paper we used the data from SDSS DR9 release \citep{ahn2012}. The 
SDSS catalogs were queried from local SQL database. The spatial
queries to the database were handled using Q3C \citep{koposov06}.

It is important to point out that the SDSS photometric pipeline was not designed to 
deal with morphologically complicated  objects like Leo P, so for 
example the extended low-surface brightness areas of such galaxies 
are often de-blended as separated sources; some of the \hii\ regions are
labeled as extended sources, while some others as point sources. 
Furthermore when searching for low surface brightness objects 
the diffraction spikes and glares from bright stars provide significant 
contamination of candidate lists. These problems complicate the search
significantly and in the end limit the automation of the process.
The main selection criteria adopted were the following:

\begin{itemize}
\item Presence of blue point sources  $-0.5<(g-r)_0<1$, $-2<(r-i)_0<0$ 
consistent with \hii\ regions. 
In order to avoid artefacts from bright
stars, we required that the 
SDSS photometric flags
for these objects would not have \textsc{SUBTRACTED} 
and \textsc{SATURATED} fields 
set\footnote{\url{https://www.sdss3.org/dr8/algorithms/flags\_detail.php}}.
 The objects having higher density of such sources within 10, 20, 30 arcseconds were ranked higher.
\item Presence of one or several of faint extended sources 
($\rm modelmag_g>20$) within 10, 20, 30 arcseconds.
\item Absence of bright ($\rm modelmag_g<18$) galaxy nearby.
\item Absence of known LEDA\footnote{Lyon-Meudon Extragalactic Database: \url{http://leda.univ-lyon1.fr/}} source or bright UCAC\footnote{USNO CCD Astrograph Catalog: \url{http://www.usno.navy.mil/USNO/astrometry/optical-IR-prod/ucac}} star nearby. 
\item High galactic latitude.
\end{itemize}

These search criteria yielded hundreds of candidates (including Leo P
and many known nearby star-forming dwarf galaxies). 
We then inspected the list, removed artefacts and all known sources within NED (NASA/IPAC Extragalactic Database), while keeping the best cases to create a final list of
$\sim 100$ low surface brightness star-forming dwarf galaxies. A small subset of these is reproduced in
Figure~1, and their most important properties
are collected in Table~\ref{tab:gals}. For simplicity,
in this paper we refer to the individual
objects as `KJ~X', where X is
the identification number and KJ corresponds to
`Koposov-James'.

To illustrate the importance of different cuts to produce a small list of candidates, 
we provide a summary of the number of different sources which ``survive" a few key stages of our selection: 
\begin{itemize}
\item The number of high latitude $|b|>30$ primary SDSS sources with $psfmag_g<21$ is  $\sim$ 58 million.
\item After applying the $g-r$ and $r-i$ colour cuts listed above, the number of objects is reduced to 
$\sim 2$ million. 
\item If we require to have two or more sources satisfying our colour-cut inside a 20$\arcsec$ aperture, the number of objects is reduced to $\sim$ 50000 or $\sim$ 8300 if we require having more than three objects.
\item After requiring not to have objects with SATURATED or SUBTRACTED photometric flags, 
those numbers reduce to
34000 and 6400 respectively. 
\end{itemize}
Further cuts, such as cross-match with LEDA, UCAC as well as bright galaxies ($modelmag_g<18$) remove the residual contamination of the list from ``big" galaxies and artefacts from bright stars, 
and reduce the list of candidates to manageable size.

Our objects were additionally cross-referenced with the catalogue of 21~cm \hi\ line sources of \citet{Haynes:2011}, consisting of 15,855 sources detected within 40\%\ of the ALFALFA survey \citep{Giovanelli:2005}.  Six of the 12 objects presented here are detected within the \hi\ survey (KJ~1, 3,7,18,22, and 32) and we utilise the \hi\ masses presented in \citet{Haynes:2011} for these galaxies in Section~\ref{sec:morph}.  The Arecibo General Catalogue (AGC) identification numbers for these galaxies are listed in Table~\ref{tab:gals}.  Furthermore, the optical photometry-based properties of KJ~7  have been reported \citet{Huang:2012b}, who classified this object as a `dwarf' according to the ALFALFA criteria.  Overall, with respect to our entire sample, we find that $\sim$20\%\ of our objects have a \hi\ detection within the ALFALFA survey.

It should also be noted that one galaxy within our sample, KJ~53, already exists within the sample of XMPs presented by \citet{Morales-Luis:2011} (object 18 in their table~1).  This overlap between the two selection techniques may result from the fact that this particular object (i) lies at the closest distance within our current sample and (ii) is the most compact in nature.  Being photometry driven, our selection criteria typically result in more distant and diffuse objects than those previously detected, although some degree of overlap is expected (and corrected for). \footnote{In this case of KJ~53, a direct method abundance had not been measured or published within the literature and as such we felt it was beneficial to keep it within the MMT sample.}

\section{Observations and data reduction}
The galaxies studied here are listed in Table~\ref{tab:gals} in order of KJ number, along with some of their general properties such as their
coordinates and redshift.  
Spectroscopic observations were obtained for each galaxy 
with the Multiple-Mirror Telescope (MMT) Blue-Channel Spectrograph on the nights of 27--28th January, 2014,  and objects were chosen on account of their visibility during these nights.
Observations were made using the medium-resolution grating 
(300 grooves~mm$^{-1}$), giving a spatial scale along the slit of 0.6\arcsec\ per pixel, a spectral range of 3500--8000~\AA, and a spectral resolution of $\sim$7~\AA (FWHM).
Table~\ref{tab:gals} also provides information on the position angle (PA) of the spectrograph slit, and exposure times of the observations.    

The processing of the spectra was performed using a custom pipeline written in Python. The reduction steps were standard. The raw data were bias subtracted and flat-fielded. The wavelength calibration was performed using the arcs observed immediately after of before each science exposure. The sky spectrum was subtracted from each science frame. The spectra of science objects were extracted by summing the flux along the slit within some aperture; optimal extraction were not possible due to the complicated structure of the objects (multiple \hii\ regions with different line ratios  and galaxy continuum). The extracted spectra from multiple exposures were combined using weighted average together with sigma-clipping to exclude
residuals from cosmic rays. The flux calibration of the spectra was performed using spectrophotometric standard stars observed throughout each night with the instrument set-up described above.
Uncertainties resulting from flux calibration were estimated to be $\sim$8--12\%, which is incorporated into the error spectrum accordingly.  This somewhat conservative error is applied in order to account for the effects of systematic uncertainties, e.g. less than photometric conditions etc.  In order to minimise these errors, each target was observed at its parallactic angle if it fell below an airmass of $\sim$ 1.1 during the observation. In the cases where multiple \hii\ regions were clearly evident in the two-dimensional spectra, extraction profiles were adjusted to obtain separate spectra of each region.  Such cases are denoted has `KJ~X.1, KJ~X.2' etc. 


\begin{figure*}
\includegraphics[]{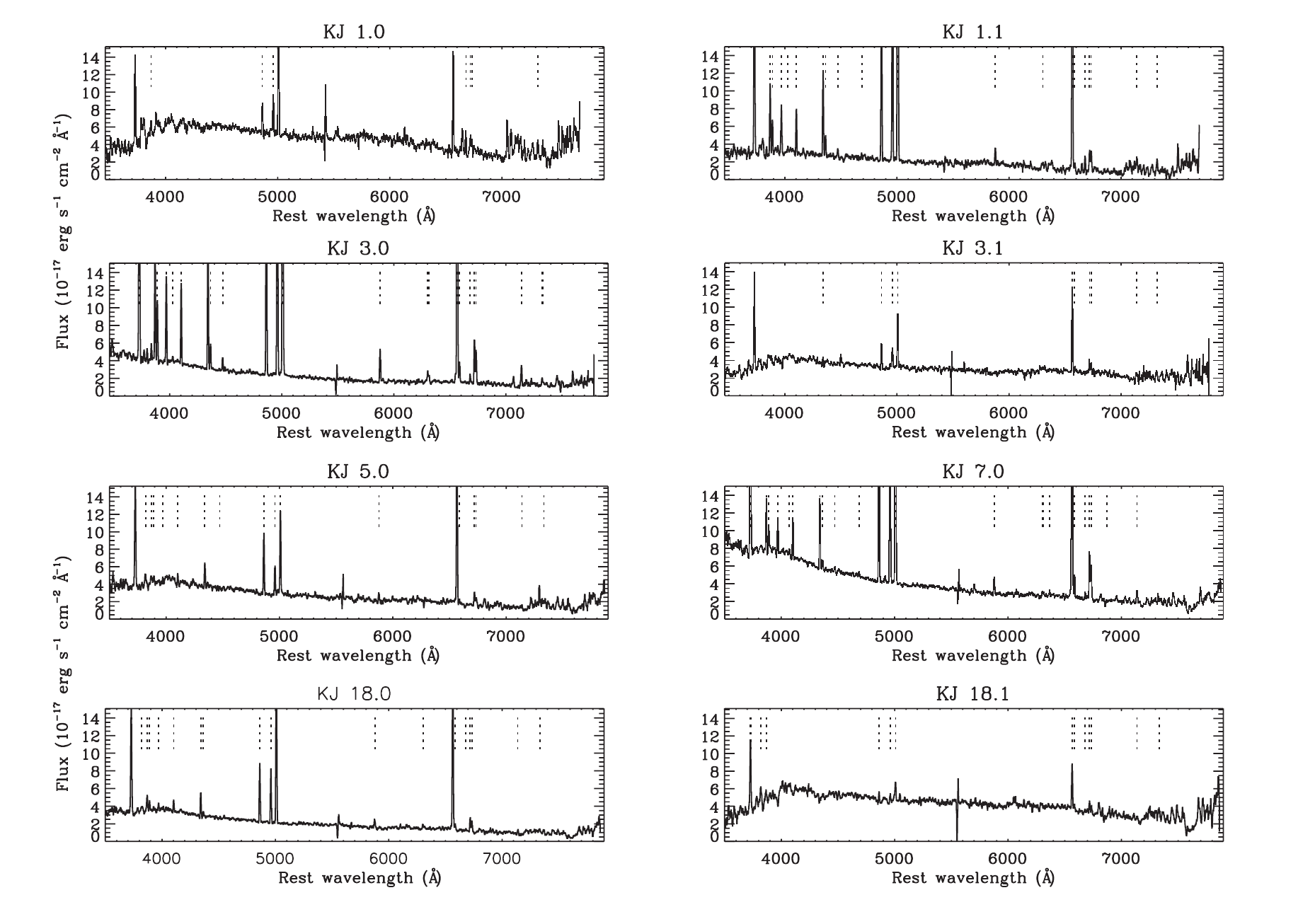}
\caption{MMT spectra of the sample of blue diffuse dwarf galaxies detailed in Table~\ref{tab:gals} and shown in Figure~\ref{fig:images}.  Wavelengths are given in the rest frame of each galaxy.  Each spectrum has been smoothed with a 5-pixel boxcar for presentation purposes. The $y$-scale has been chosen to show the weakest emission lines. 
Dashed-lines indicate emission lines detected in each spectrum, as listed in Table~\ref{tab:fluxes_all}.  Solid red lines indicate non-detected \hb.}
\end{figure*}
\begin{figure*}
\setcounter{figure}{1}
\includegraphics[width=1.0\textwidth]{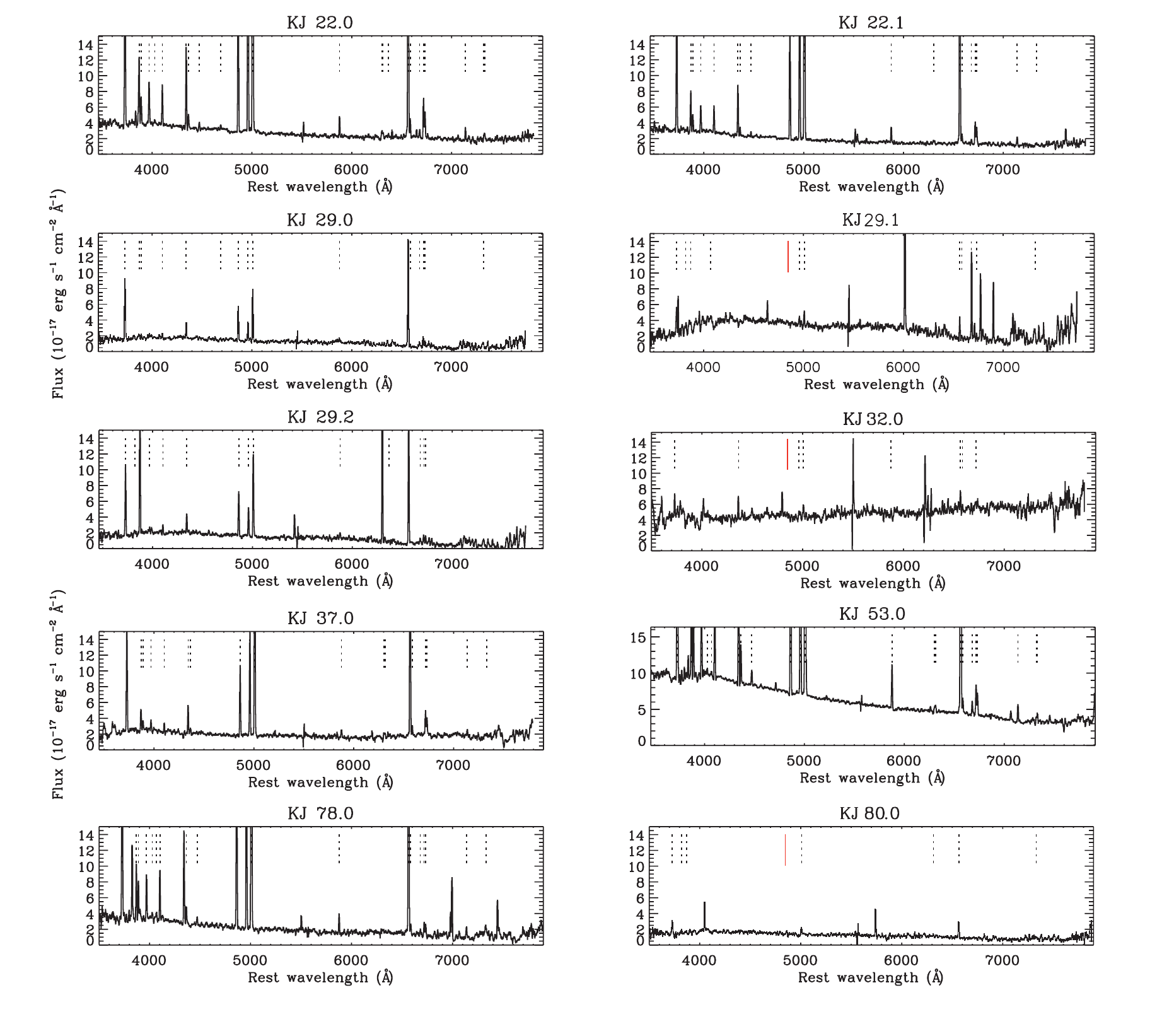}
\caption{-- continued.}
\label{fig:spec}
\end{figure*}
\section{The spectra}
Figure~\ref{fig:spec} shows a complete, flux-calibrated MMT spectrum for each galaxy within our sample (as detailed in Table~\ref{tab:gals}).  Each spectrum exhibits all the strong emission lines typical of star-forming galaxies, superimposed on a blue continuum.  In some of the brighter objects, we also observe the typically fainter emission lines such as the \hei\ optical recombination lines and high ionisation lines of \fsiii\ and \fariii.  We also show three objects (KJ~29.1, 32.0, and 80.0) whose spectra were particularly faint and where \hb\ was not detected at 3-$\sigma$ significance level.  As such, we have decided not to include these three objects in the analysis described within this paper, and simply show them here for completeness purposes.

For each spectrum, we model the emission lines with a single Gaussian component (multiple Gaussian components were not required), to deduce their redshifts and total integrated line flux of each emission line.  The resultant redshifts are given in Table~\ref{tab:gals}, while the detected lines and their integrated line fluxes are listed in Table~\ref{tab:fluxes_all}.  
Redshifts were found to range from $z\sim 0.001$ to 0.023.
Line flux measurements and their uncertainties were obtained following the methodology outlined in \citet{James:2014a}, which employs a MCMC (Markov chain Monte Carlo) method to fit all of the observed lines within each spectrum simultaneously (i.e., requiring the redshifts and line widths to be the same for all the lines fitted).  

Table~\ref{tab:fluxes_all} lists all lines that were found to be above the $3 \sigma$ detection limit.  
We also list the $3 \sigma$ upper limits of undetected lines that we deem necessary for the abundance calculations described in Section~\ref{sec:abund} (e.g. \foiii~$\lambda$4363 and \fnii~$\lambda$6585).  The upper-limit measurements were obtained by integrating the error spectrum over the model profile width determined from the high S/N emission lines.   For galaxies in which underlying stellar absorption was visible, i.e. those with EW(\ha)$<100$~\AA, Balmer line fluxes were corrected accordingly (in the case of \hb, equivalent widths of the underlying profiles were found to be 1--3\AA).

Table~\ref{tab:fluxes_all} also lists the de-reddened fluxes along with the reddening values used to make this correction. Reddening was estimated from the observed
F(\ha)/F(\hb) and F(\hg)/F(\hb) ratios (combined in a 3:1 ratio, respectively), adopting the case B recombination ratios (appropriate for gas with $T=10,000$~K and \eld=100~cm$^{-3}$) and the LMC extinction curve \citep{Fitzpatrick:1999}.  In cases where \hg\ was not detected, only the F(\ha)/F(\hb) ratio was employed. For two of our targets (Objects KJ~1.1 and KJ~3) we adopt $E(B-V)=0.0$ (i.e. zero reddening) because the Balmer decrements were slightly below the theoretical case B ratio; this is most likely due to the uncertainties in the flux calibration.

\begin{table*}
\caption{Ionic and elemental abundances for the BDD galaxies presented within this paper, derived from the emission line measurements given in Table~\ref{tab:fluxes_all}.  Electron density and temperatures are also listed.}
\begin{center}
\begin{footnotesize}
\begin{tabular}{|l|cccccccc|}
\hline
 ID & \eld(\foii,\fsii ) & \elt(\foiii)  & \elt(\foii) & \op/\hp & \opp /\hp  & 12+log(O/H) & log(N/O) & [O/H]$^a$ \\
 & (cm$^{-3}$)  & (K) & (K) & ($\times10^{-5}$) & ($\times10^{-5}$) & & & \\ 
\hline
KJ~1.0& 
         590$^{+        2020}_{-         290}$&
$<$       17470&
$<$       14650&
$>$   3.69&
$>$   1.76&
$>$   7.74&
$>$  -1.74&
$>$  -0.95\\
KJ~1.1& 
         720$^{+        1530}_{-         400}$&
       15380$\pm$        1960&
       12860$\pm$        1370&
   2.01$\pm$   0.87&
   3.18$\pm$   1.06&
   7.72$\pm$   0.16&
  -1.51$\pm$   0.21&
  -0.97$\pm$   0.16\\
KJ~3.0& 
         110$^{+         520}_{-          60}$&
       12930$\pm$        2100&
       11360$\pm$        1290&
   4.27$\pm$   2.58&
   5.56$\pm$   3.12&
   7.99$\pm$   0.23&
  -1.60$\pm$   0.29&
  -0.70$\pm$   0.24\\
KJ~3.1& 
         460$^{+        1700}_{-         230}$&
$<$       19720&
$<$       16250&
$>$   3.07&
$>$   0.79&
$>$   7.59&
$>$  -1.23&
$>$  -1.10\\
KJ~5.0& 
         180$^{+         700}_{-         130}$&
$<$       15430&
$<$       12830&
$>$   3.77&
$>$   0.86&
$>$   7.67&
$>$  -1.69&
$>$  -1.02\\
KJ~7.0& 
        1500$^{+        1180}_{-         660}$&
       12880$\pm$        1540&
       11330$\pm$         950&
   6.32$\pm$   2.58&
   3.60$\pm$   1.33&
   8.00$\pm$   0.16&
  -1.47$\pm$   0.20&
  -0.69$\pm$   0.17\\

KJ~18.0& 
         690$^{+        2060}_{-         340}$&
       16280$\pm$        2770&
       13540$\pm$        1990&
   3.01$\pm$   1.62&
   1.62$\pm$   0.63&
   7.67$\pm$   0.20&
  -1.55$\pm$   0.22&
  -1.02$\pm$   0.20\\
KJ~18.1& 
        2480$^{+          50}_{-        1180}$&
$<$       54020&
$<$       44710&
$>$   0.61&
$>$   0.14&
$>$   6.87&
$>$  -1.08&
$>$  -1.82\\
KJ~22.0& 
         230$^{+         540}_{-         130}$&
       12920$\pm$        1390&
       11330$\pm$         860&
   4.26$\pm$   1.62&
   5.33$\pm$   1.83&
   7.98$\pm$   0.15&
  -1.49$\pm$   0.19&
  -0.71$\pm$   0.16\\
KJ~22.1& 
         120$^{+         420}_{-          70}$&
       13550$\pm$        1580&
       11690$\pm$        1020&
   3.87$\pm$   1.57&
   4.18$\pm$   1.46&
   7.91$\pm$   0.16&
  -1.59$\pm$   0.20&
  -0.78$\pm$   0.17\\
KJ~29.0& 
         320$^{+        1090}_{-         250}$&
$<$       15960&
$<$       13150&
$>$   2.91&
$>$   0.91&
$>$   7.58&
$>$  -1.60&
$>$  -1.11\\
KJ~29.2& 
        1140$^{+        1750}_{-         660}$&
$<$       13360&
$<$       11380&
$>$   5.02&
$>$   1.72&
$>$   7.83&
$>$  -2.10&
$>$  -0.86\\
KJ~37.0& 
         650$^{+        1950}_{-         290}$&
       14930$\pm$        2090&
       12720$\pm$        1400&
   3.88$\pm$   1.77&
   2.92$\pm$   1.06&
   7.83$\pm$   0.17&
  -1.49$\pm$   0.21&
  -0.86$\pm$   0.18\\
KJ~53.0& 
         300$^{+         720}_{-         160}$&
       16890$\pm$        2290&
       13770$\pm$        1750&
   1.03$\pm$   0.47&
   1.80$\pm$   0.57&
   7.45$\pm$   0.15&
  -1.52$\pm$   0.19&
  -1.24$\pm$   0.16\\

KJ~78.0& 
         380$^{+         760}_{-         260}$&
       15020$\pm$        1930&
       12420$\pm$        1380&
   1.64$\pm$   0.76&
   2.53$\pm$   0.86&
   7.62$\pm$   0.16&
  -1.39$\pm$   0.20&
  -1.07$\pm$   0.17\\
\hline
\end{tabular}

\end{footnotesize}
\end{center}\label{tab:abunds}
$^a$ [O/H]$\equiv$log(O/H)$_{gal}-$log(O/H)$_{\odot}$, where log(O/H)$_{\odot}$ is from the solar composition value recommended by \citet{Asplund:2009}.
\end{table*}

\section{Chemical abundances}\label{sec:abund}
In order to calculate chemical abundances, we adopted the `direct method' - i.e. a `direct' measurement of the abundance of an element relative to \hp\ based on the electron temperature (\elt) and density (\eld) of the gas and the extinction-corrected line fluxes.  Ionic abundances are then converted into elemental abundances by adding up the different stages of ionisation.  It is this latter point that prevents us from measuring several different elemental abundances and we instead limit our calculations 
to O/H and N/O.\footnote{Whilst we regularly detect lines necessary to calculate \Sp, \neppp, and \arppp, the spectra are not deep enough to detect  higher ionisation stages and we would therefore be forced to employ ionisation correction factors (ICFs).  ICFs can often introduce large uncertainties \citep{Barlow:1994} and we therefore refrain from using them within this study.}


When calculating the total elemental abundance, each of the ionic abundances are derived using the electron temperatures corresponding to the ionisation `zone' in which that ion dominates.  For example, in order to calculate \opp/\hp, we calculate \elt(\foiii) using the ratio of \foiii\ lines to an auroral line (e.g. \foiii($\lambda5007+\lambda4959$)/$\lambda4363$).  In order to sample the lower ionisation zone of the \hii\ region, where \op/\hp\ resides, we require the auroral lines of \foii.  However, these are often weak and not detected, as is the case here.  While \elt(\foii) can be calculated using empirical methods, a considerable amount of uncertainty exists due to outdated atomic data (energy levels, transition probabilities, and collision strengths).  We therefore follow the methodology described in \citet{Nicholls:2014} to calculate total oxygen abundances using equations derived from first principles and the most up-to-date atomic data.  

The method itself is iterative, firstly calculating \opp/\hp\ using \elt(\foiii) from the ratio described above.  The \opp\ is treated as the total oxygen abundance and used to calculate \elt(\foii) via an empirical relationship.  The \op/\hp\ abundance can then be calculated using this value, and the \foii~$\lambda\lambda$3727, 3729 doublet.  The total O/H abundance is then recalculated (\op$+$\opp) and the process repeated.  The calculation converges within five iterations and involves case B emissivities and collision strengths for \foii\ and \foiii\ based on their respective temperatures.  \citet{Nicholls:2014} also includes a means of calculating log(N/O) by combining the methodology described above with the empirical formulae of \citet{Izotov:2006}, which account for the temperature dependencies of \foii\ and \fnii.

For each galaxy within our sample, we use the extinction-corrected line fluxes listed in Table~\ref{tab:fluxes_all} 
and the method described above to calculate \elt(\foiii), \elt(\foii), \op/\hp, \opp/\hp, 12+log(O/H), and log(N/O) and list them in Table~\ref{tab:abunds}.  In addition, we provide the mean \eld\ calculated using the density sensitive \foii~$\lambda\lambda$3727, 29 and \fsii~$\lambda\lambda$6716, 32 doublets, and adopting \elt(\foii).  Uncertainties in both 
values of \elt\ correspond to uncertainties in the relevant extinction-corrected fluxes, and are then propagated into the 
errors in the abundance values.

Accurate measurements of oxygen abundances (i.e. from detections of
\foiii~$\lambda 4363$) were obtained for nine out of 15 spectra
(eight out of 10 objects).
For those spectra where \foiii~$\lambda$4363 was considered a non-detection, we list the oxygen abundances as lower limits.  Out of those for which values of \elt\ were obtainable, oxygen abundances were found to be in the range $7.45 < 12+\log{\rm (O/H)} <8.0$, with two out of eight objects, KJ~53 and KJ~78, falling within the XMP classification 
($12+\log{\rm (O/H)}<7.65$).  


Nitrogen-to-oxygen ratios were found to lie between 
$-1.60 < \log {\rm (N/O)} < -1.39$, and are typical values for the metallicity range of the sample \citep[see e.g.][their figure 11]{Lopez-Sanchez:2010}.   This ratio, and its relation to metallicity, is discussed in more detail in Section~\ref{sec:NO}.

With regards to homogeneity between multiple \hii\ regions within the same system (i.e. KJ~1,18, 22, 29, and 3), only KJ~22 has oxygen abundances that were measured with certainty in both regions and they are in agreement (within the uncertainties).  The remaining candidates, where one is measured as a limit, are also consistent with having homogeneous oxygen abundances between the \hii\ regions.  Nebular abundances trace the youngest population and the chemical composition of the star-forming material, and so the observed chemical homogeneity may suggest that the gas within these objects is well mixed and/or star-formation began at a similar time throughout.  

\section{Emission-line properties}\label{sec:props}

\begin{table*}
\begin{footnotesize}
\caption{Various properties of the sample of blue diffuse dwarf galaxies: \ha\ luminosity, star-formation rate, ionisation parameter ($U$) and the age of the current star-forming population determined from the equivalent width of \hb.  A detailed description of each property can be found in Section~\ref{sec:props}.}
\begin{center}
\begin{tabular}{|l|cccc|}
\hline
ID & $L$(\ha)  & SFR(\ha) & log(U)  & Age (EW(\ha,\hb)) \\
 & ($\times10^{38}$ erg\,s$^{-1}$) & ($\times10^{-2}$ \Msol \,yr$^{-1}$)  & & (Myr) \\
\hline
  
KJ 1.0$^{\star}$ &  31.48$\pm$  3.59 &   2.49$\pm$  0.28         & -3.01 $\pm$ 0.10                             & $<$  11.81 \\  
KJ 1.1 &  98.20$\pm$  9.85 &   7.76$\pm$  0.78         & -2.58 $\pm$ 0.11                                       &  6.63 $\pm$  1.32 \\  
KJ 3.0 &  56.24$\pm$  8.96 &   4.44$\pm$  0.71         & -2.50 $\pm$ 0.18                                       &  4.27 $\pm$  0.67 \\  
KJ 3.1$^{\star}$ &   6.70$\pm$  0.68 &   0.53$\pm$  0.05         & -3.25 $\pm$ 0.07                             & $<$ 12.19  1.30 \\  
KJ 5.0$^{\star}$ &   1.08$\pm$  0.12 &   0.09$\pm$  0.01         & -3.19 $\pm$ 0.07                             &  $<$ 9.85   0.84 \\  
KJ 7.0 &   0.99$\pm$  0.10 &   0.08$\pm$  0.01         & -2.78 $\pm$ 0.13                                       &  6.09 $\pm$  1.52 \\  
KJ 18.0 &   1.18$\pm$  0.13 &   0.09$\pm$  0.01       & -2.96 $\pm$ 0.08                                        &  9.55 $\pm$  0.91 \\  
KJ 18.1$^{\star\dagger}$ &   0.39$\pm$  0.05 &   0.03$\pm$  0.00        & -3.47 $\pm$ 0.08                      & $<$ 15.70  \\  
KJ 22.0 &  32.19$\pm$  3.40 &   2.54$\pm$  0.27        & -2.52 $\pm$ 0.11                                       &  4.96 $\pm$  1.09 \\  
KJ 22.1 &  20.03$\pm$  2.12 &   1.58$\pm$  0.17        & -2.61 $\pm$ 0.10                                       &  6.06 $\pm$  1.47 \\  
    
KJ 29.0$^{\star}$ &  19.92$\pm$  2.07 &   1.57$\pm$  0.16        & -3.13 $\pm$ 0.08                             &  $<$ 9.96 \\  
KJ 29.2$^{\star}$ &  31.34$\pm$  3.43 &   2.48$\pm$  0.27        & -3.00 $\pm$ 0.09                             &  $<$ 8.30  \\  
KJ 37.0 &  28.82$\pm$  3.62 &   2.28$\pm$  0.29        & -2.79 $\pm$ 0.10                                       &  7.69 $\pm$  1.50 \\  
    
KJ 53.0 &   0.47$\pm$  0.05 &   0.04$\pm$  0.00        & -2.65 $\pm$ 0.09                                       &  7.23 $\pm$  1.58 \\  

KJ 78.0 &   0.43$\pm$  0.04 &   0.03$\pm$  0.00        & -2.61 $\pm$ 0.09                                       &  6.65 $\pm$  1.74 \\  
  
\hline
\end{tabular}
\label{tab:SFR}
\end{center}
\end{footnotesize}
$^\star$ Object has lower limit oxygen abundance\\
$^\dagger$ Calculated using $Z=0.1$~\Zsol\ (the minimum metallicity allowed within the $U$-parameter relation of 

\citet{Kewley:2002}).
\end{table*}

\subsection{Star-formation rate and L(\ha)}
We calculate the \ha\ luminosity for each galaxy using the extinction-corrected \ha\ line fluxes (Table~\ref{tab:fluxes_all}) and the distances given in Table~\ref{tab:gals}.  
\ha\ luminosities are in the range  
$37.6 < \log(L_{{\rm H}\alpha}) < 40.0$\,erg~s$^{-1}$, 
at the upper end of the range in luminosities seen in dwarf irregular galaxies \citep[e.g.][]{vanZee:2000b} and in the lower range of luminosities seen in BCDs \citep[e.g.][]{GildePaz:2003}.  
\ha\ equivalent widths (EW) were also found to be typical of star-forming dwarf galaxies, 
falling within $20\lesssim {\rm EW( H}\alpha)  \lesssim300$~\AA.
 
The \ha\ luminosities were converted into a star-formation rates using the calibration of \citet{Kennicutt:1998}.  Both $L$(\ha) and SFR(\ha) are listed in Table~\ref{tab:SFR}.  It can be seen that our current subsample of BDDs displays a rather large range in current SFRs, 
from $0.03 \times 10^{-2}$ to $7.8 \times 10^{-2}$\,\Msol~yr$^{-1}$, 
a range which is consistent with the range seen in local quiescent 
star-forming galaxies (i.e. $ < 0.1$\,\Msol~yr$^{-1}$), and lower than in starburst galaxies ($\sim 0.1$--1~\Msol~yr$^{-1}$).  

Overall, XMP galaxies are often found to have starbursting SFRs and as such are sometimes referred to as `emission line XMPs'.  For these cases, \citet{Ekta:2010} associate the combination of high SFR and low-metallicity 
with an infall of \hi\ gas from the outskirts of the galaxy or accretion of metal-poor gas.  
However, our subsample of BDDs shows SFRs which are more reminiscent of `quiescent XMPs'  -- a class of XMPs 
which became apparent with the discovery of Leo~P, 
whose low luminosity 
\citep[$L_{{\rm H}\alpha}=6.2\times10^{36}$\,erg\,s$^{-1}$,][]{Rhode:2013} corresponds to a SFR of just $5\times10^{-5}$\,\Msol~yr$^{-1}$. 
\citet{Skillman:2013} suggest that in comparison to their emission-line counterparts, quiescent XMPs may instead have formed through normal evolution of very low-mass galaxies.   Indeed, inline with Leo~P, 
objects with particularly low SFRs (i.e. $<1\times10^{-3}$\Msol~yr$^{-1}$) 
in our sample are also those with the lowest metallicities.  

Using SFRs we can separate the currently observed BDD sample into two classes; those which are actively star-forming (SFR$>$0.01~\Msol~yr$^{-1}$) and those with more quiescent SFRs (i.e. low but non-zero SFRs).  On the one hand we have low-luminosity star-forming galaxies, while on the other we have (relatively) high-luminosity quiescent galaxies.  For the former, it is their diffuse nature that has allowed them to pass below the detection threshold in previous emission-line surveys.


\begin{figure*}
\includegraphics[scale=0.5,angle=90]{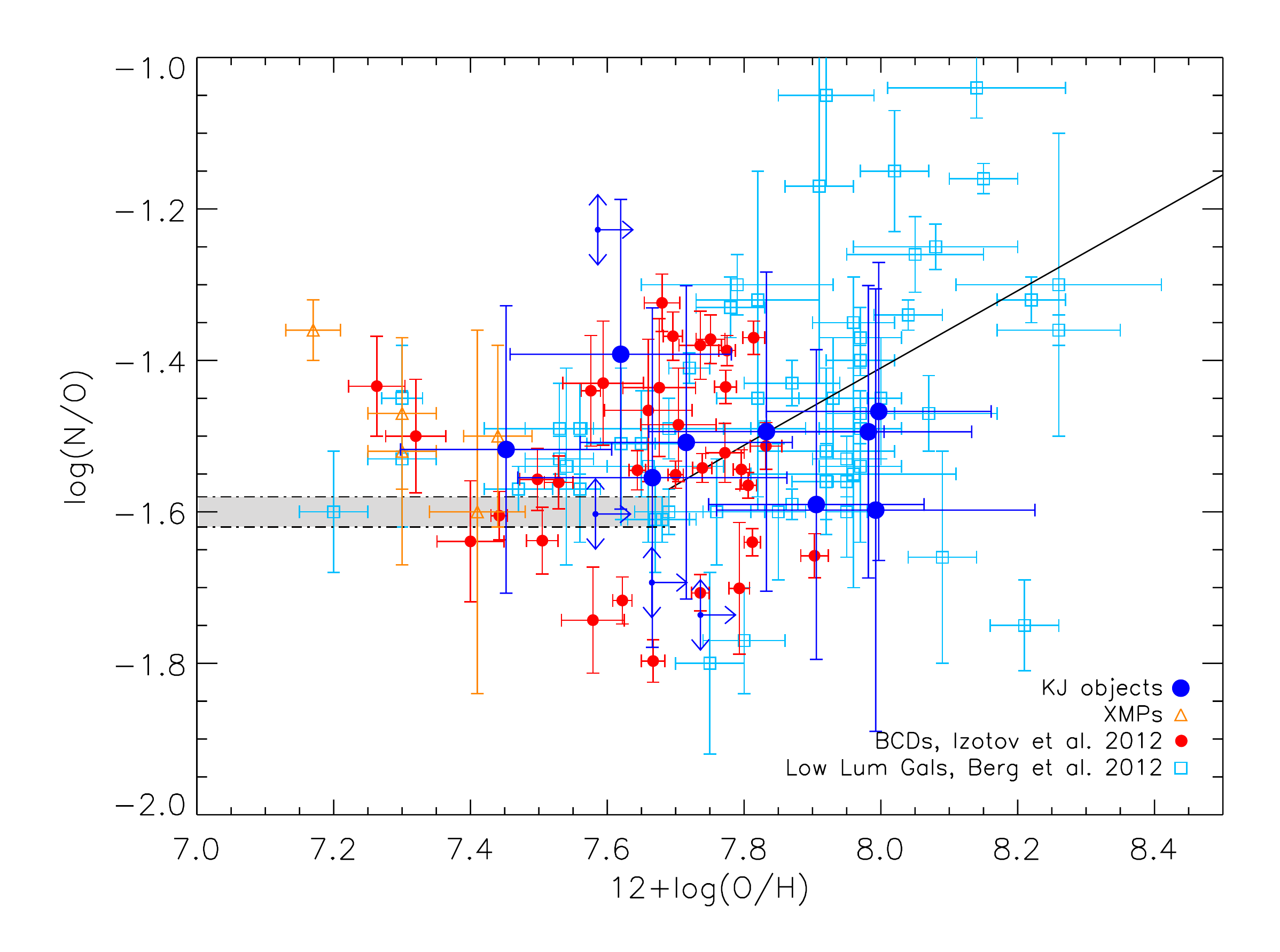}
\caption{The relationship between metallicity and nitrogen-to-oxygen ratio
in the low-metallicity regime. We show the results from our sample of BDD galaxies (i.e. KJ~objects listed in Table~\ref{tab:gals}, blue circles) and for comparison we plot the values measured in a sample of XMPs (see text for details), the recently compiled sample of BCDs by \citet{Izotov:2012} and the low-luminosity dwarf galaxies of \citet{Berg:2012}.  The black solid line shows the relationship 
between log(N/O) and log(O/H) in the metallicity range
$12+\log{\rm (O/H)} > 7.7$ derived by \citet{Berg:2012}
for low-luminosity galaxies, while the grey shaded region is the narrow N/O plateau proposed for XMPs by \citet{Izotov:1999}. } 
\label{fig:met_no}
\end{figure*}

\subsection{Ionisation strength}
In order to assess the strength of the ionising field within our subset of BDDs, we  calculate the ionisation parameter, $U$.  This dimensionless parameter represents the ratio of the density of ionising photons per unit volume to the particle (atom plus ion) number density.  The $U$-parameter is often quoted as $q$, where $U=q/c$, $c$ being the speed of light.  We estimate the value of $U$ for each galaxy using the iterative equations of \citet{Kewley:2002} together with the oxygen abundances listed in Table~\ref{tab:abunds} and the \foiii~$\lambda$5007/\foii~$\lambda\lambda$3727, 3729 ratios.  Adopting the metallicity of each galaxy, we interpolate between the relevant equations of \citet{Kewley:2002}.  The resultant $U$-parameters are given in Table~\ref{tab:SFR}, and range between 
$-3.47< \log(U) < -2.50$ ($7.0 < \log(q) < 7.9$).  These are somewhat typical values for mid-to-low metallicity star-forming galaxies \citep[see e.g.][]{Dopita:2013}.

\subsection{Current age of ionising stellar population}
We investigate the current age of the star-forming population via the luminosities of hydrogen recombination lines. \ha\ and \hb, in particular, provide an estimate of the ionising flux present, assuming a radiation-bounded nebula \citep{Schaerer:1998}.  We therefore estimate the age of the latest star-forming episodes in each object via the Balmer line equivalent width and its metallicity, and compare them with those predicted by the spectral synthesis code, Starburst99 \citep{Leitherer:2010}.  Models were run at the full range of available metallicities (0.05, 0.2, 0.4, and 1~\Zsol), along with assumptions of an instantaneous burst, a Salpeter IMF, `standard' Geneva tracks and Paudrach-Hillier atmospheres.  Only models with zero-rotation were employed due to the limited range in metallicities currently available for models with rotation.   We interpolate between the models at the metallicity of each galaxy (Table~\ref{tab:abunds}), and assess the age of the current stellar population according to its EW(\hb) and EW(\ha).  We list the mean and difference of these two ages in Table~\ref{tab:SFR}.

Ages were found to range between 4.3 and 9.6~Myr, 
with a mean age of 6.6~$\pm$1.5~Myr (this does not account for galaxies for which lower limit oxygen abundances are available, as only upper limits can be calculated in these cases).  
Such young ages suggest that these BDDs are within the earlier stages of star-formation or have recently undergone a burst of star-formation.  At this particular age, stellar populations are expected to be going through their Wolf-Rayet (WR) phase.  While there are hints of WR-signatures, i.e. the `WR bump' at $\sim$4690~\AA\, in several of the objects (KJ~1.1, 3, 7, 22 and 29), the S/N of the spectra is not sufficient to quantify the population age
more precisely.   It is interesting to note, however, that WR features are not typically seen in XMP galaxies, even when the starburst is young \citep{Shirazi:2012}.

\section{Discussion}\label{sec:disc}
In order to assess the role that blue diffuse dwarfs play within the local population of galaxies we will compare the properties discussed in Sections~\ref{sec:abund} and \ref{sec:props} with those of other dwarf galaxies. We do note, however, that since only a small number of BDDs have been studied to date, such comparisons may suffer from low-number statistics.

We are particularly interested in how our current subset of BDDs compare to other low-metallicity dwarf galaxies and have therefore assembled a comparison sample consisting of the following: blue compact dwarfs (BCDs) from  \citet{Izotov:2012}; XMP galaxies [Leo~P \citep{Skillman:2013}, LeoA \citep{VanZee:2006}, UGCA292 \citep{VanZee:2000}, DDO68 \citep{Pustilnik:2005}, SBS1129+576 \citep{Ekta:2006,Guseva:2003}, SBS1129+577 \citep{Ekta:2006}, J2104-0035 and UGC772 \citep{Ekta:2008}, UM133 and SDSSJ011914 \citep{Ekta:2010}, HS2134 \citep{Pustilnik:2006} the sample of \citet{Morales-Luis:2011}]; low-luminosity dwarf galaxies of \citet{Berg:2012}.  We also include nearby dwarf galaxies from the volume-limited survey by \citet{Karachentsev:2013}.  In order to avoid discrepancies between metallicity measurements, we only include galaxies for which a direct measurement of the metallicity has been
reported.  

\subsection{N/O vs. Oxygen abundance}\label{sec:NO}
A great deal of discussion regarding the relationship between the oxygen abundance and the N/O ratio exists in the literature, as such a 
relationship can be used to assess the chemical evolutionary age of a galaxy  \citep[see e.g. ][and references therein]{Skillman:2003,Lopez-Sanchez:2010b}. 
It was first thought that at extremely low metallicities N/O 
is constant with O/H, forming a narrow plateau which 
\citet{Izotov:1999} attributed to XMPs currently undergoing their \textit{first} burst of star-formation (i.e. ages $\lesssim40$~Myr) and nitrogen undergoing primary production by high-mass stars.  However, as the number of low-metallicity objects has increased, the reality of such a narrow range
of values of N/O has come to be questioned. 
For example, \citet{VanZee:2006} reported a large scatter in N/O 
at low metallicities, which had been anticipated by \citet{Garnett:1990}.
The scatter may be due to the time delay between the release 
of oxygen from massive stars and primary nitrogen
from low- and intermediate-mass stars as proposed by
\citet{Vila:1993} \citep[see also][]{Pettini:2008}, or perhaps an infall of metal-poor gas in some galaxies \citep{Amorin:2012b},
but a full interpretation is still far from being
established \citep{Skillman:2013, Zafar:2014}.
At higher metallicities, the effects of secondary production 
are seen as N/O increases linearly with oxygen abundance.

As can be seen from Fig.~\ref{fig:met_no}, the values of N/O we have
found in our sample of BDDs are comparable to those measured 
in other low-metallicity galaxies. 
The limited accuracy of our abundance
determinations and limited sample size preclude firm conclusions
concerning the reality of the scatter at the lowest 
metallicities. Rather we stress the future potential 
of BDD galaxies for addressing, and possibly resolving,
this issue as they provide a firmer handle on the relationship in the low-metallicity regime.

\begin{figure*}
\includegraphics[scale=0.5,angle=90]{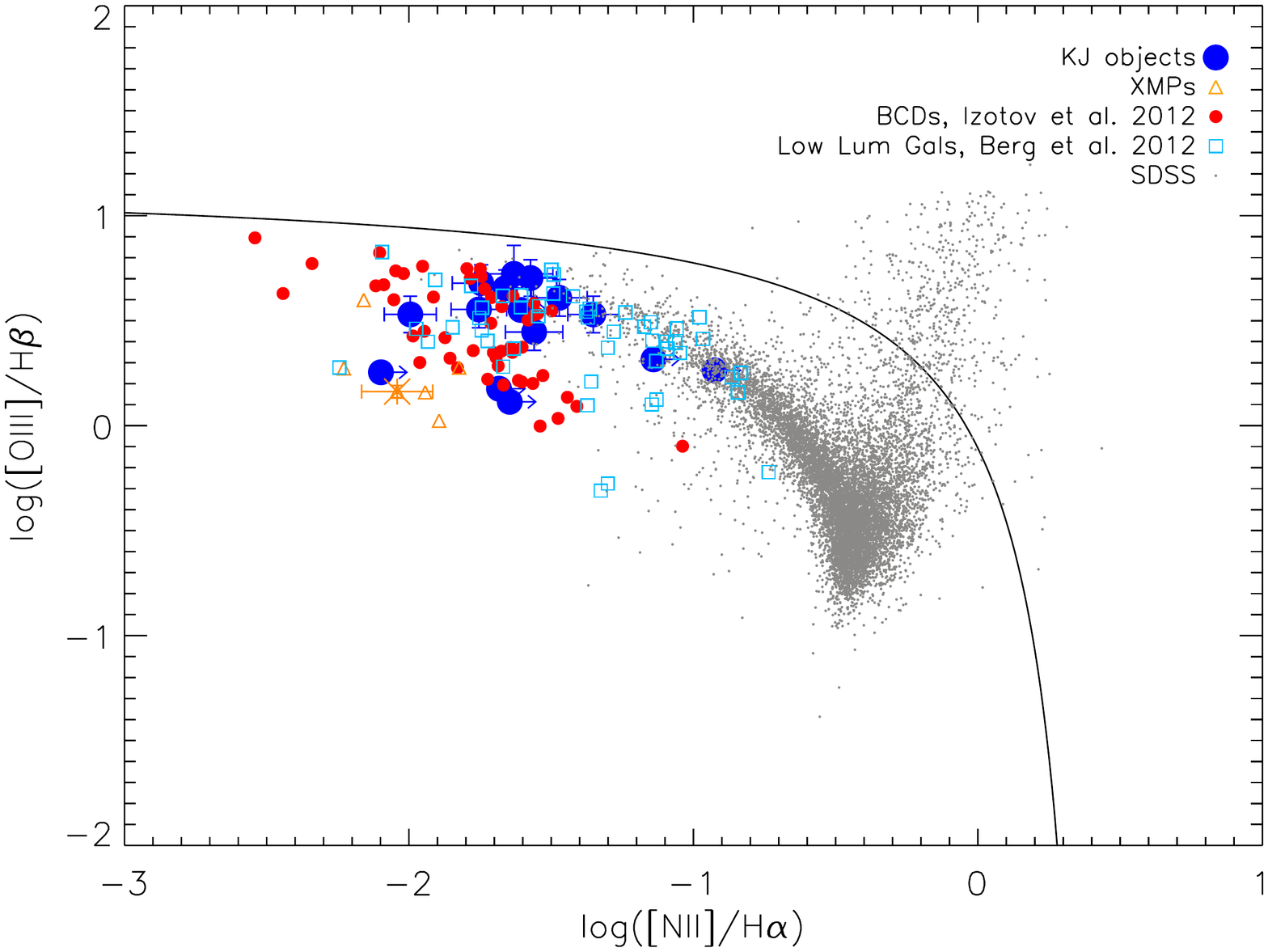}
\vspace{-1cm}
\caption{An emission line diagnostic diagram, 
showing \foiii/\hb\, vs. \fnii/\ha, traditionally used to separate galaxies according to their metallicity, strength, and hardness of their ionising radiation.  In black we show the SDSS emission line galaxies (selected according to the criteria of \citet{Brinchmann:2008}), which form a branch of star-forming galaxies on the left, and Seyferts/LINERs on the right.  The BDD galaxies (i.e. KJ~objects listed in Table~\ref{tab:gals})  tightly cluster towards the top part of the star-forming branch, suggesting they comprise of low-metallicity, highly-ionised gas.   For comparison, we also plot a selection of XMPs, BCDs and low-luminosity dwarf galaxies (see text for details).  The black solid line represents the `maximum starburst line' of \citep{Kewley:2001}.}
\label{fig:BPT}
\end{figure*}
\subsection{Emission line diagnostics}
In Figure~\ref{fig:BPT} we place the BDDs on the well known emission line diagnostic diagram, \fnii/\ha\ vs. \foiii/\hb.  This diagram, first utilised by \citet{Baldwin:1981} and \citet{Veilleux:1987}, separates galaxies with regards to their metallicity, strength of their ionising radiation, and hardness of the ionising radiation.  In order to demonstrate this, we plot the emission line ratios of the SDSS galaxies 
in Figure~\ref{fig:BPT}, which form the well know `seagull' shape -- a branch of star-forming galaxies on the left and a branch of Seyfert/LINERs on the right.  It can be seen from Fig.~\ref{fig:BPT} that BCDs, due to their starburst and low-metallicity environments, tend to lie on the top-left branch of the star-forming galaxies.  XMPs, on the other hand, tend to fall slightly below the BCDs, suggesting somewhat weaker ionising radiation perhaps due to a more quiescent star-forming environment.  In comparison to the other dwarf galaxies, the BDDs within our existing sample are quite tightly clustered and all lie towards the top of the star-forming branch and within the upper-range of \fnii/\ha\ displayed by BCDs.  Their distribution within the diagram suggests that they are objects consisting of low-metallicity, highly ionised gas.  However,
in general BDDs do not seem to overlap with XMP galaxies.

\begin{table}
\caption{Photometric properties of the BDD sample.}
\begin{center}
\begin{tabular}{lcccc}
\hline\hline
ID   &       M$_B$       &         $\mu_B$   & log($M_\star$) & log($M_{HI}$) \\
 & (mag)  & (mag arcsec$^{-2}$) & (\Msol)  & (\Msol) \\
\hline

KJ 1   &   -17.98  &  24.1 & 9.04 &9.39  \\
KJ 3   &   -17.14  &  24.4  & 8.90 & 9.40 \\
KJ 5   &   -13.12  &  24.5  & 8.43 & ---\\
KJ 7   &   -13.28  &  24.3  & 7.48 & 7.60 \\
KJ 18  &  -15.49  &  23.2   &8.43 & 8.44\\
KJ 22  &  -15.92  &  23.5  & 8.32 & 8.77 \\
KJ 29  &  -17.31  &  24.7  & 8.79 & --- \\
KJ 37  &  -16.24 &  24.4  & 8.69 & --- \\
KJ 53  &  -11.17  &  22.6   & 6.45 & ---\\
KJ 78  &  -10.72  &  24.6  &6.52 & ---\\
\hline
\end{tabular}
\end{center}
\label{tab:phot}
\end{table}%
\begin{figure*}
\includegraphics[scale=1]{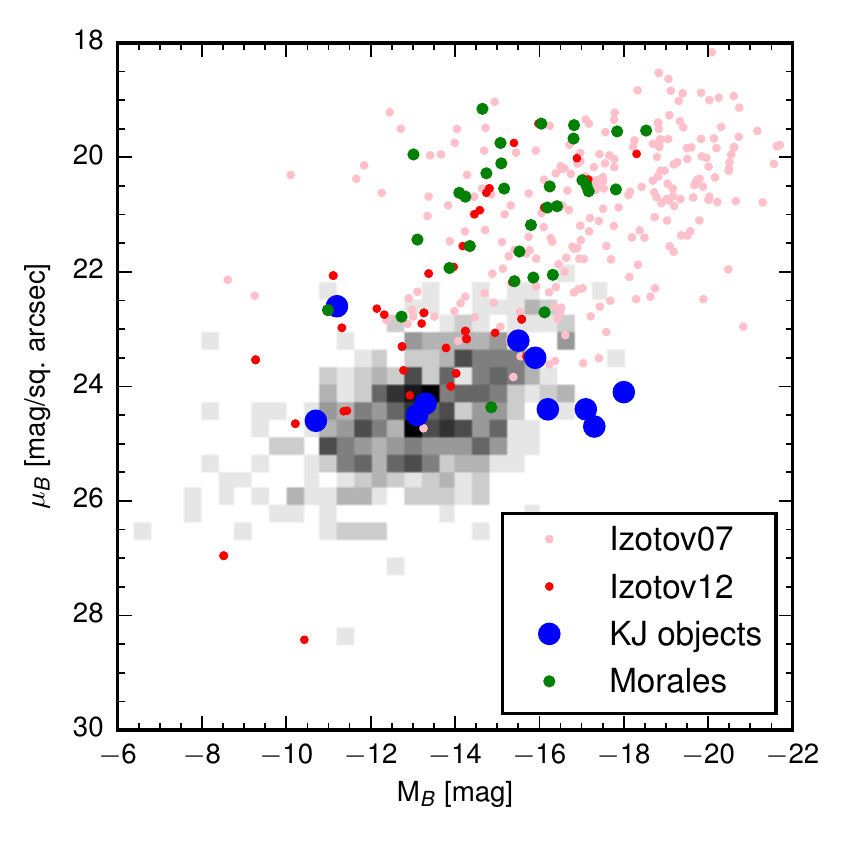}
\caption{Absolute $B$-band magnitude versus effective $B$-band surface brightness for the BDD galaxies within our current sample (i.e. KJ~objects listed in Table~\ref{tab:gals}) .  We also overplot the SDSS-selected low-metallicity \hii\ regions \citep{Izotov:2007c} and XMP galaxies \citep{Morales-Luis:2011,Izotov:2012}, along with nearby dwarf galaxies from the volume-limited sample by \citet{Karachentsev:2013} shown in grey-scale.}
\label{fig:Mb_mu}
\end{figure*}

\section{Morphological Properties and Stellar Masses}\label{sec:morph}
Whilst our sample of BDDs are clearly dwarf galaxies, deciding which particular class of dwarf galaxies they belong to is somewhat difficult.  
Taking into account their irregular morphology, low metallicities and current level of star formation, BDDs are most naturally placed within the dwarf irregular galaxies (dIrr) category.  The different types of dwarf  irregular galaxies have not been precisely defined, and classification is often based on their morphology and/or current rates of star formation.

Within the dIrr category we also have blue compact dwarf galaxies -- an extreme form of dIrr where star-formation occurs in a few compact regions, and at an intense rate.  BCDs, by definition, are compact -- a term which refers more to the size of the high surface brightness component than the total optical size.  Peak surface brightnesses of BCDs are significantly higher than most other dIrrs, 
with BCDs having $\mu_{\rm B,peak}< 22$~mag\,arcsec$^{-2}$ \citep{GildePaz:2003}.  
This is mostly due to the recent star-forming episodes within the BCDs, whereas the star-formation in dIrrs is less recent and less active, 
causing them to be fainter and redder.  Unfortunately, due to the highly irregular distribution of light throughout the BDD galaxies, we are 
at present unable to quantify the peak surface brightness of our
objects; this is something which we intend to achieve with better quality photometric observations. 

We can however assess their effective surface brightness $\mu_B$. To compute this parameter we calculate the total B-band magnitude and the half light radius of each galaxy from their $g$ and $r$-band growth curves obtained from the SDSS images. To convert the $g$-band and $r$-band magnitudes into the $B$-band we use the relations from \citet{Lupton:2005}.
Both the absolute $B$-band magnitudes and effective surface brightnesses for the BDDs within the sample are listed in Table~\ref{tab:phot}. 

With the aim of putting the morphological properties of BDDs into context, in Figure~\ref{fig:Mb_mu} we compare their magnitudes and surface brightnesses against those of other dwarf galaxies observed in the SDSS survey.  By only using other SDSS dwarfs we can ensure  consistency between the calculations $M_B$ and $\mu_B$, which we derive using the method described above.  Our comparative sample consist of the XMPs of \citet{Morales-Luis:2011}, low-metallicity emission line galaxies of \citet{Izotov:2012}, low metallicity \hii\ regions of \citet{Izotov:2007c}, and the volume-limited sample of \citet{Karachentsev:2013}.  It can be seen that BDDs lie within the fainter end of low-metallicity dwarf galaxies, and overlap with the local dwarf irregular galaxies.  Their position within the diagram suggests that BDDs could be the more distant analogues of local dwarf galaxies, which are currently undergoing or recently finished a burst of star-formation.  However, rather than the star-formation occurring in compactly distributed \hii\ regions, it occurs in a more diffuse manner and this leads to lower values of $\mu_B$.  As such, this separates BDDs from BCDs and XMPs and places them within the middle of the distribution, suggesting that these systems could be a link between quiescent dIrrs and starbursting BCDs. 

Using the $g-$ and $r-$band colours, we can also make a crude estimation of the stellar mass, $M_\star$, within the BDDs using the stellar mass-to-light ratio relation $(M/L)_i=0.222+0.864\times(g-r)$ from \citet{Bell:2003}.  Stellar masses are found to range between 
${\rm M}_\star \sim 0.03 \times 10^{8} {\rm M}_\odot$ and
$\sim 11 \times 10^{8} {\rm M}_\odot$ (see last column
of Table~\ref{tab:phot}). This range encompasses the stellar masses of
both dIrrs \citep[see e.g.][]{Lee:2006} and BCDs \citep[see e.g.][]{Zhao:2013}.  It should be noted however that without IR images, such estimates are extremely uncertain \citep[e.g.][]{GildePaz:2003,Zhao:2013} and as such we refrain from including them in any further analysis of the population.

For comparison, in Table~\ref{tab:phot} we also list the \hi\ masses for those galaxies detected within the ALFALFA survey, as presented in \citet{Haynes:2011}.  Overall, it appears that the BDD's baryonic mass is dominated by atomic gas, rather than their stars.  This was also found to be the case for over half of the ALFALFA dwarfs presented in \citet{Huang:2012b}.  With regards to Leo~P, $\log (M_{HI}/M_\odot)\sim$6, the BDDs are 2--3 orders of magnitude more massive in \hi.



\begin{figure*}
\includegraphics[scale=0.5,angle=90]{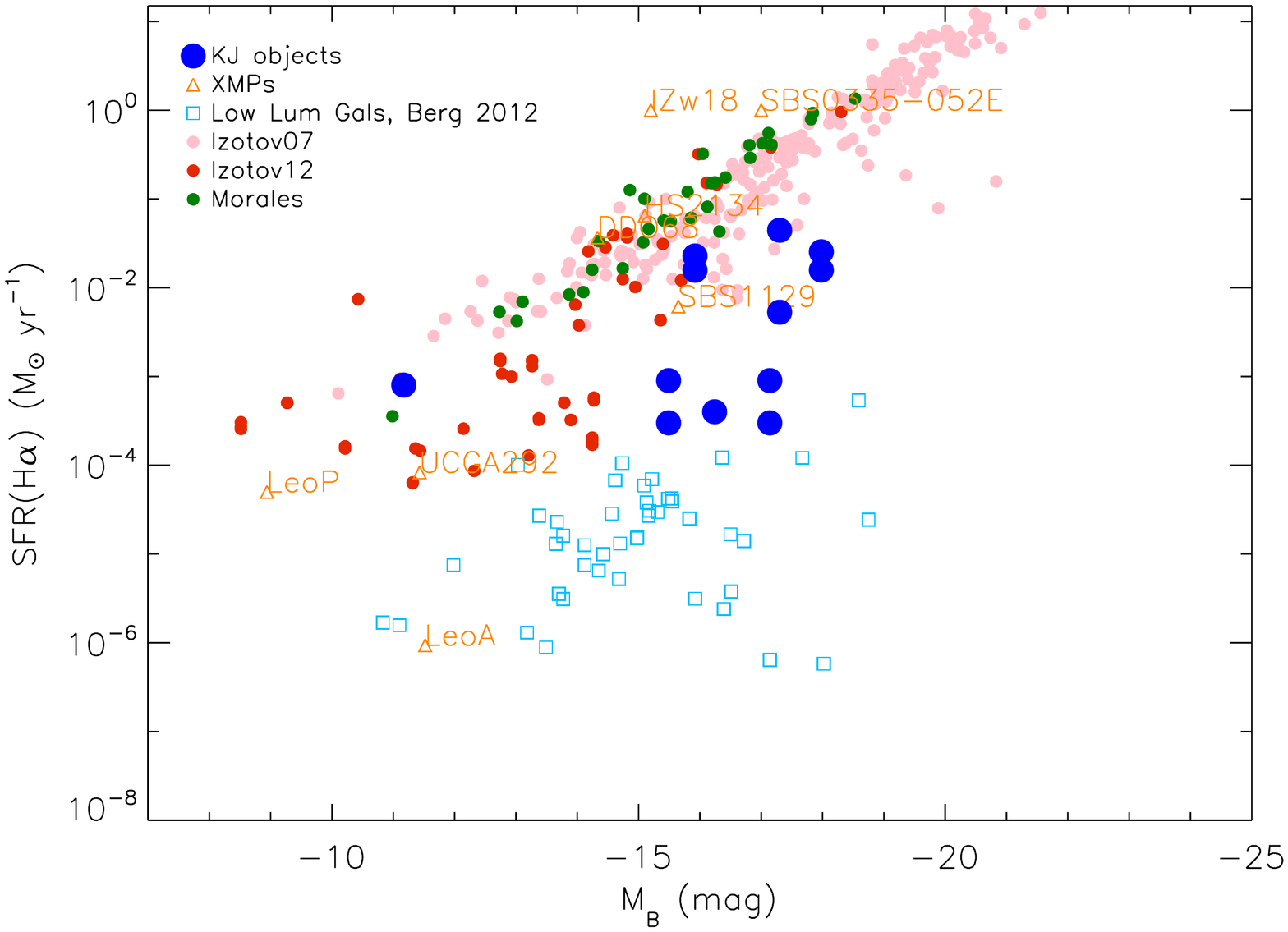}
\vspace{-1cm}
\caption{The relationship between absolute $B$-band magnitude ($M_B$) and star-formation rate for our sample of blue diffuse dwarf galaxies (i.e. KJ~objects listed in Table~\ref{tab:gals}) .  We also overplot the comparison sample described in Section~\ref{sec:disc}.}
\label{fig:Mb_SFR}
\end{figure*}

\section{What are BDDs?}
Overall, our current subset of BDD galaxies appear to be metal-poor, low surface brightness objects in the nearby universe who are forming stars at rates that range from quiescent to semi-active.  Whilst their current stellar populations are very young ($<10$~Myr) we cannot rule out the possibility of an older, underlying population, as found in all BCDs (and XMPs) close enough for us to allow resolved stellar populations \citep[e.g.][]{Aloisi:2007,Papaderos:2008}, and deep HST imaging of nearby dIrrs (e.g. Leo~A, \citet{Cole:2007}; IC~1613, \citet{Skillman:2014}; DDO~210, \citet{Cole:2014}).  While each of these properties make them very similar to other dwarf galaxies currently studied, their combination of non-zero SFR and diffuse morphology separate them somewhat.  In Figure~\ref{fig:Mb_SFR} we plot the current SFR (derived from extinction-corrected \ha\ emission) against $M_B$ for BDDs, low-luminosity galaxies, a selection of XMPs, BCDs, and individual \hii\ regions in low-metallicity galaxies.  This plot is comparable to the popular SFR vs. mass diagram used in studies of high-redshift galaxies.   In SFR-luminosity space, it can be seen that BDD galaxies occupy a region between low-luminosity quiescent galaxies and bright, star bursting BCD galaxies.  BDDs do align with the individual \hii\ regions of low-metallicity galaxies observed by \citet{Izotov:2012}, which were selected from SDSS-DR7 on the basis of having weak \foiii\, and \fnii\ emission lines relative to \hb.  Although, interestingly, most of low-metallicity host-galaxies detected by \citet{Izotov:2012} were already known in the literature - possibly because these systems have relatively higher surface brightnesses (Figure~\ref{fig:Mb_mu}).

On the one hand, BDDs have emission line properties that agree with various BCD criteria \citep[e.g.][and references therein]{GildePaz:2003}.  On the other hand, their morphological properties are more in-line with dIrrs (Figure~\ref{fig:Mb_mu}). It should be noted, however, that we base such comparisons purely on the range of properties observed within the small subset we have studied to date. Even so, with regards to their SFRs, our current sample of BDDs could be considered as a mixed bag of quiescent and semi-actively star-forming galaxies or, in other words, a mixture of dIrrs and low-luminosity BCDs.  In comparison to BCDs, dIrrs are currently forming stars at an extremely low rate which will for the majority of time remain constant and then every so often increase, and during this burst a dIrr may be classed as a BCD.  However, if a starburst episode occurs within a BDD, their diffusely distributed \hii\ regions would prevent it from displaying the compact characteristic of a BCD.  This may in part be due to the method by which the star-formation has been triggered within these systems.  In BCDs, star-formation is typically triggered by an inflow of \hi\ gas \citep[e.g.][]{Gordon:1981, Zhao:2013,Verbeke:2014} and/or interaction or merging event \citep[e.g.][]{Pustilnik:2001,Bekki:2008}, causing star-formation to occur in concentrated \hii\ regions (see \citet{Sanchez-Almeida:2014} for an extensive review of this topic).  In BDDs, however, we have star-formation occurring in a less concentrated, haphazard distribution across each system.
 
\begin{figure*}
\includegraphics[scale=0.5,angle=90]{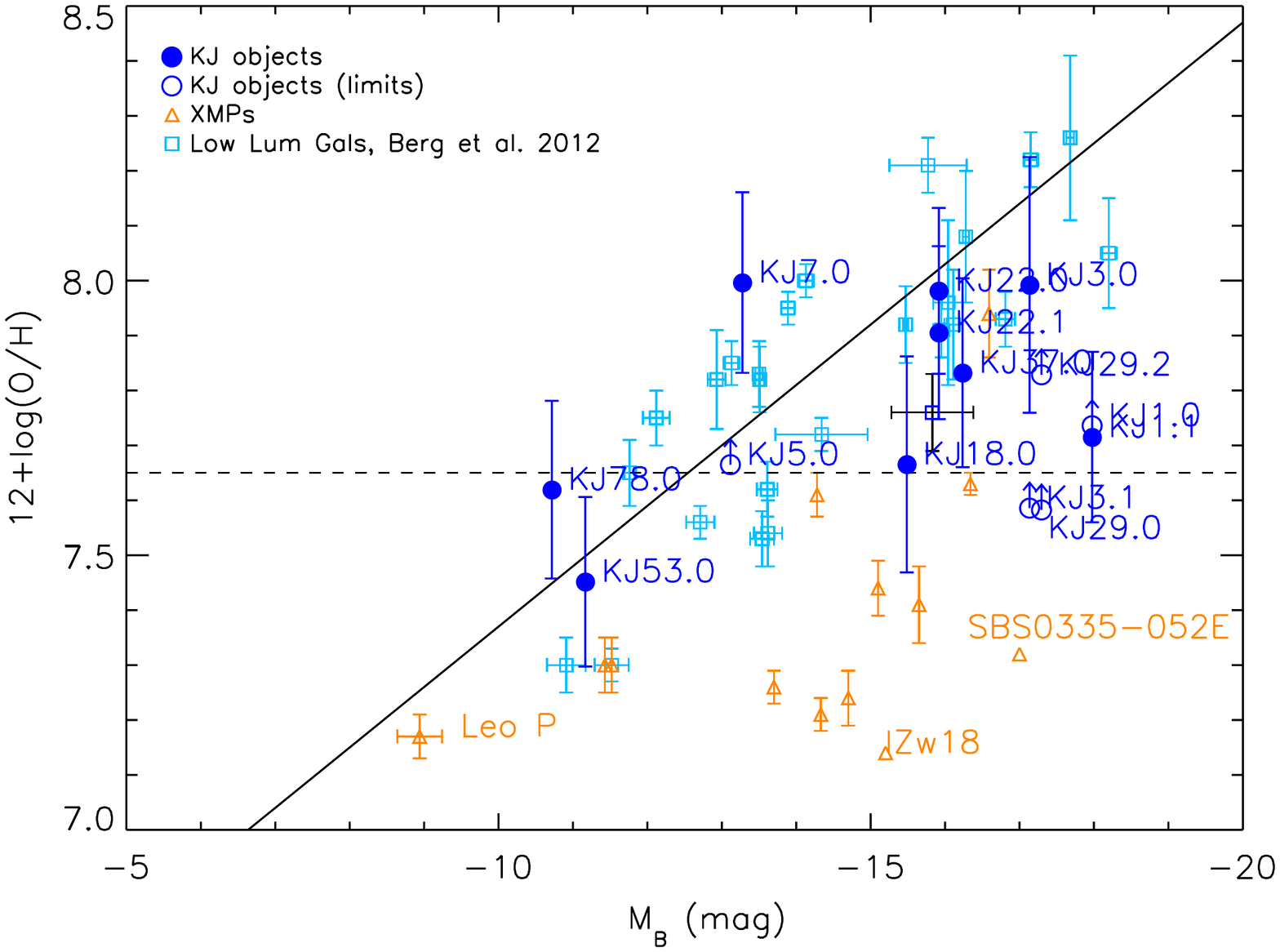}
\vspace{-1cm}
\caption{The relationship between oxygen abundance and absolute $B$-band magnitude ($M_B$)  for our sample of blue diffuse dwarf galaxies (i.e. KJ~objects listed in Table~\ref{tab:gals}).  We also overplot the comparison sample described in Section~\ref{sec:disc}.}
\label{fig:mb_met}
\end{figure*}
The inflow of metal-poor gas into BCDs and other XMP galaxies has often been used to explain why these galaxies do not follow the $L-Z$ relationship seen in local star-forming dwarf galaxies.  We illustrate this in Figure~\ref{fig:mb_met}, where we plot the $B$-band magnitudes and metallicities of low luminosity galaxies and the $L-Z$ relationship of \citet{Berg:2012}, along with XMP galaxies of \citet{Morales-Luis:2011}, and other low-metallicity record holders (I~Zw~18, SBS~0335-052E, and Leo~P).  Pristine gas flowing into XMP and BCD galaxies not only depletes the ISM chemical abundances but also enables increased star-forming activity in and around the area of impact.  In support of this, many of these outliers are known to have disturbed kinematics due to interaction and/or infall \citep{VanZee:1998a,Ekta:2006, Ekta:2008,Ekta:2009}.  By contrast, Leo~P and the BDD galaxies are in agreement with the $L-Z$ relation for local dwarf galaxies.  Leo~P is both extremely quiescent with regards to star-formation and shows no signs of interaction and/or infall  \citep{Bernstein:2014}, and is therefore thought have formed through normal evolution of low mass galaxies \citep{Skillman:2013}.  Whilst BDDs are more actively star-forming than Leo~P and other low luminosity galaxies (Figure~\ref{fig:Mb_SFR}), the distribution of star-forming regions is still reminiscent of dIrrs, where star-formation occurs in randomly distributed \hii\ regions.  However, in order to explore the possibility of inflowing metal-poor gas, we ultimately require integral field unit data.  Such observations can provide spatially resolved kinematics, velocity dispersions, and metallicities, and the combination of these three parameters can be used to recognise the presence of inflows or outflows \citep[see e.g.][]{James:2013a}.
    
From the information currently available to us we propose that BDDs belong to a sub-class of dwarf irregular galaxies which are forming stars more actively than typical dIrrs.  Their distribution of star-forming regions (i.e. randomly scattered amongst a diffuse blue continuum, Fig.~1), low-metallicities, and emission line ratios all suggest they are the diffuse counterparts to BCDs, where star-formation occurs in smaller, more isolated \hii\ regions, and under more `normal' circumstances.  It is this latter physical characteristic that has so-far prevented them from being detected in emission-line surveys.  Our sample of objects shows that by focussing on a galaxy's morphological properties, we can uncover alternatives to the bright, compact objects systematically found using traditional emission-line-based search techniques.  Our results complement those of blind \hi\ surveys, which are also largely successful in uncovering faint, blue cloud-like objects, and can reach down to very low mass objects \citep[i.e. $\log M_{HI}/M_\odot<7.7$][]{Huang:2012b}.  Whilst their metallicities are not as low as their model equivalent, Leo~P, the fact that 2 out of 8 objects can be classified as XMPs suggests that our morphology-based search technique may well be successful in filling in the missing low-metallicity dwarf galaxies.  Upon reflection of our current sample, the range in metallicities and SFRs observed in BDDs suggest that (i) Leo~P may well be an outlier with respect to these properties or (ii) our criteria are not efficient in selecting objects with metallicities as low as Leo~P.  However, both of these propositions are a consequence of distance (the two XMP galaxies uncovered via our search criteria are also the closest i.e. within 7~Mpc) and without a a specific volume-limited survey, we cannot fully assess the density/frequency of such extreme objects.
 
\section{Conclusions}

With the aim of uncovering rare XMP galaxies in the local Universe, we undertook a morphological-based search on SDSS imaging data, using the structural properties of the recently discovered XMP galaxy, Leo~P as a guide.  Our unique search algorithm  yielded $\sim$100 blue, diffuse galaxies containing one or more \hii\ regions embedded in low-surface brightness emission; we have termed these objects  blue diffuse dwarf galaxies (BDDs).  

In this first report, we have presented preliminary results from ongoing optical spectroscopic follow-up observations of the BDD galaxies using the MMT spectrograph.  The 12 galaxies discussed here lie at distances of $\sim$5--120~Mpc and $\sim$40\arcsec\ across.  We were able to extract separate spectra from multiple \hii\ regions within five of the galaxies.  All of the spectra showed bright emission lines typical of star-forming galaxies, with the exception of three objects (KJ~29.1, 32.0 and 80.0), which were excluded from further analysis.  The final spectroscopic sample available for analysis consisted of a total of 15 spectra from 10 galaxies.  The main results are as follows:
\begin{enumerate}
\item{Electron temperature measurements were achieved for nine out of 15 spectra, amounting to eight out of 10 galaxies, and for these we were able to derive `direct' measurements of oxygen abundances.  The systems were found to have metallicities in the range $7.45 < 12+\log({\rm O/H})<8.0$.
Two of the galaxies fall within the common definition of 
extremely metal-poor galaxies (XMPs) with 
$12+\log({\rm O/H}) < 7.65$ (less than 1/10 solar).}
\item{Nitrogen-to-oxygen ratios were found to be in the range
 $-1.60< \log{\rm (N/O)} < -1.39$; such ratios are typical for the
 above values of oxygen abundance.}
\item{The galaxies we have observed exhibit  
current rates of star-formation (derived from \ha\ luminosities) ranging between $0.03\times10^{-2}$ and $7.8 \times10^{-2}$\,\Msol~yr$^{-1}$.}
\item{The ages of the stellar populations responsible for the star formation we see are between 4.3 and 9.6~Myr, with a mean age of $\sim 7$~Myr.}
\item{Due to their diffuse morphology, peak surface brightness estimates were not obtainable from SDSS images.  However, we do estimate the effective surface brightness, $\mu_B$, 
to lie between 23 and 25~mag~arcsec$^{-2}$.  Absolute $B$-band magnitues, $M_B$ are between $\sim-11$ and $\sim-18$~mag. }
\end{enumerate}
Overall, from the information gained on this sub-sample, BDDs appear to be a population of dwarf galaxies that lie in between the quiescent dIrrs and star-bursting blue compact galaxies.  Both their surface brightnesses and star-formation rates are intermediate between these two dwarf categories, suggesting that BDDs may represent more `normal' metal-poor star-forming systems.   However, in order to quantify and classify these galaxies fully, we will require additional data at a range of wavelengths.  In particular, spatially resolved observations would be critical in obtaining kinematical information, whilst IR imaging would enable us to measure their mass.

If we return to the original motivation of this work, i.e. whether XMP galaxies are more readily detected if we focus on their diffuse `Leo\,P-type' morphology, we have shown that while morphology-based searches have the ability to uncover objects that are \textit{similar} to Leo\,P, objects with metallicities as low as Leo\,P appear to be extremely rare.  We defer further discussion of this point to a future
paper, once we have observed spectroscopically a larger fraction
of our sample of BDDs.

We have demonstrated that searching for XMP galaxies using only their emission line properties may be preferentially biased towards finding those currently or recently undergoing a burst of star-formation, or with star-formation occurring in compactly distributed \hii\, regions.
  By instead focussing on their optical morphological properties, we have uncovered metal-poor galaxies in which star-formation is ongoing in isolated \hii\ regions distributed within a diffuse continuum - a structural property which can lead to such galaxies being overlooked by most emission-line searches.

\section*{Acknowledgments}
Observations reported here were obtained at the MMT Observatory, a joint facility of the University of Arizona and the Smithsonian Institution. We are grateful to the University of Arizona Observatory time assignment committee who awarded time to this programme, and thank the MMT telescope operators and staff for technical support.  We are indebted to Alis Deason for providing the first-look spectroscopic observations of a subset of our sample that were used to support our initial MMT proposal. It is a pleasure to acknowledge Rob Kennicutt for useful discussions concerning dwarf galaxy populations, Matt Auger who helped with various aspects of the data analysis, and Evan Skillman for insightful comments on an early version of this work.  We sincerely thank the reviewer of the paper, Jorge Sanchez Almeida, whose helpful comments and suggestions greatly improved the paper.   EWO is partially supported by NSF grant AST1313006.  The research leading to these results has received funding from the European Research Council under the European Union's Seventh Framework Programme (FP/2007-2013)/ERC Grant Agreement no. 308024. SK acknowledges financial support from the ERC.  This research has made use of the NASA/IPAC Extragalactic Database (NED) which is operated by the Jet Propulsion Laboratory, California Institute of Technology, under contract with the National Aeronautics and Space Administration.
\bibliographystyle{mn2e}
\bibliography{references}
\clearpage

\clearpage

\appendix
\begin{landscape}
\section{Observed Fluxes and Line Intensities}
Spectroscopic line fluxes and de-reddened line intensities (both relative to \hb$=100$) as measured from the spectra shown in Figure~\ref{fig:spec}. 
\begin{center}
\begin{footnotesize}
\begin{tabular}{|l|cc|cc|cc|cc|}
\hline
 & \multicolumn{2}{c|}{ KJ~1.0} & \multicolumn{2}{c|}{ KJ~1.1 }& \multicolumn{2}{c|}{ KJ~3.0} & \multicolumn{2}{c|}{ KJ~3.1} \\
\hline\hline
\foii\       3727.10 & 
     170$\pm$     27 &      199$\pm$     33 & 
    70.0$\pm$   10.2 &     70.0$\pm$   10.2 & 
    58.3$\pm$   12.9 &     58.3$\pm$   12.9 & 
     212$\pm$     32 &      222$\pm$     33 \\ 
\foii\       3729.86 & 
     140$\pm$     23 &      164$\pm$     27 & 
    57.3$\pm$    8.3 &     57.3$\pm$    8.3 & 
     101$\pm$     22 &      101$\pm$     22 & 
     193$\pm$     29 &      202$\pm$     30 \\ 
\hei\        3820.84 & 
--- & --- & 
--- & --- & 
--- & --- & 
--- & --- \\ 
\fneiii\       3870.16 & 
    41.9$\pm$    8.5 &     47.9$\pm$    9.7 & 
    33.5$\pm$    4.8 &     33.5$\pm$    4.8 & 
    33.1$\pm$    7.3 &     33.1$\pm$    7.3 & 
--- & --- \\ 
 \hei+H8       3889.75 & 
--- & --- & 
    19.1$\pm$    2.8 &     19.1$\pm$    2.8 & 
    18.1$\pm$    4.0 &     18.1$\pm$    4.0 & 
--- & --- \\ 
\heii\       3969.55 & 
--- & --- & 
    24.7$\pm$    3.6 &     24.7$\pm$    3.6 & 
    24.5$\pm$    5.4 &     24.5$\pm$    5.4 & 
--- & --- \\ 
\hei\        4027.49 & 
--- & --- & 
    2.31$\pm$   0.74 &     2.31$\pm$   0.74 & 
    1.51$\pm$   0.42 &     1.51$\pm$   0.42 & 
--- & --- \\ 
\fsii\       4069.75 & 
--- & --- & 
--- & --- & 
--- & --- & 
--- & --- \\ 
\hd\       4102.89 & 
--- & --- & 
    24.3$\pm$    3.5 &     24.3$\pm$    3.5 & 
    24.9$\pm$    5.5 &     24.9$\pm$    5.5 & 
--- & --- \\ 
\hg\       4341.68 & 
--- & --- & 
    43.9$\pm$    6.3 &     43.9$\pm$    6.3 & 
    46.1$\pm$   10.2 &     46.1$\pm$   10.2 & 
    54.8$\pm$    9.1 &     55.9$\pm$    9.3 \\ 
\foiii\        4364.44 & 
$<$   9.33  & $<$   9.95  & 
    10.4$\pm$    1.6 &     10.4$\pm$    1.6 & 
    7.95$\pm$   1.78 &     7.95$\pm$   1.78 & 
$<$   6.95  & $<$   7.09  \\ 
\hei\        4472.93 & 
--- & --- & 
    4.68$\pm$   0.91 &     4.68$\pm$   0.91 & 
    3.81$\pm$   0.88 &     3.81$\pm$   0.88 & 
--- & --- \\ 
\feiii\        4659.35 & 
--- & --- & 
--- & --- & 
--- & --- & 
--- & --- \\ 
\heii\        4687.02 & 
--- & --- & 
    2.12$\pm$   0.64 &     2.12$\pm$   0.64 & 
--- & --- & 
--- & --- \\ 
\ariv\        4741.49 & 
--- & --- & 
--- & --- & 
--- & --- & 
--- & --- \\ 
\hb\        4862.72 & 
     100$\pm$     15 &      100$\pm$     15 & 
     100$\pm$     14 &      100$\pm$     14 & 
     100$\pm$     22 &      100$\pm$     22 & 
     100$\pm$     15 &      100$\pm$     15 \\ 
\foiii\        4960.29 & 
     118$\pm$     18 &      117$\pm$     18 & 
     154$\pm$     21 &      154$\pm$     21 & 
     174$\pm$     38 &      174$\pm$     38 & 
    68.9$\pm$   10.7 &     68.7$\pm$   10.7 \\ 
\foiii\       5008.24 & 
     362$\pm$     54 &      355$\pm$     54 & 
     477$\pm$     67 &      477$\pm$     67 & 
     528$\pm$    116 &      528$\pm$    116 & 
     207$\pm$     30 &      206$\pm$     30 \\ 
\fnii\       5756.24 & 
--- & --- & 
--- & --- & 
--- & --- & 
--- & --- \\ 
\hei\        5877.59 & 
--- & --- & 
    8.46$\pm$   1.35 &     8.46$\pm$   1.35 & 
    10.0$\pm$    2.2 &     10.0$\pm$    2.2 & 
--- & --- \\ 
\foi\        6302.05 & 
--- & --- & 
    2.04$\pm$   0.61 &     2.04$\pm$   0.61 & 
    2.77$\pm$   0.65 &     2.77$\pm$   0.65 & 
--- & --- \\ 
\fsiii\       6313.80 & 
--- & --- & 
--- & --- & 
    1.51$\pm$   0.39 &     1.51$\pm$   0.39 & 
--- & --- \\ 
\foi\       6365.54 & 
--- & --- & 
--- & --- & 
--- & --- & 
--- & --- \\ 
\ha\       6564.61 & 
     335$\pm$     50 &      286$\pm$     43 & 
     265$\pm$     37 &      265$\pm$     37 & 
     256$\pm$     56 &      256$\pm$     56 & 
     346$\pm$     51 &      329$\pm$     48 \\ 
\fnii\        6585.20 & 
$<$   8.29  & $<$   7.06  & 
    4.76$\pm$   0.89 &     4.76$\pm$   0.89 & 
    6.00$\pm$   1.35 &     6.00$\pm$   1.35 & 
    25.1$\pm$    5.2 &     23.9$\pm$    4.9 \\ 
\hei\        6679.99 & 
    40.1$\pm$    8.8 &     33.9$\pm$    7.5 & 
    7.84$\pm$   1.41 &     7.84$\pm$   1.41 & 
    2.61$\pm$   0.61 &     2.61$\pm$   0.61 & 
--- & --- \\ 
\fsii\       6718.29 & 
    49.3$\pm$    9.1 &     41.6$\pm$    7.7 & 
    10.3$\pm$    1.6 &     10.3$\pm$    1.6 & 
    12.6$\pm$    2.8 &     12.6$\pm$    2.8 & 
    55.1$\pm$    9.8 &     52.4$\pm$    9.3 \\ 
\fsii\       6732.67 & 
    29.1$\pm$    7.3 &     24.5$\pm$    6.1 & 
    8.98$\pm$   1.44 &     8.98$\pm$   1.44 & 
    10.0$\pm$    2.2 &     10.0$\pm$    2.2 & 
    40.3$\pm$    7.1 &     38.3$\pm$    6.8 \\ 
\fneiii\        6870.16 & 
--- & --- & 
--- & --- & 
--- & --- & 
--- & --- \\ 
\fariii\       7137.80 & 
--- & --- & 
    2.23$\pm$   0.71 &     2.23$\pm$   0.71 & 
    6.20$\pm$   1.40 &     6.20$\pm$   1.40 & 
    14.3$\pm$    5.1 &     13.5$\pm$    4.8 \\ 
\foii\       7321.47 & 
    52.5$\pm$   10.4 &     42.6$\pm$    8.6 & 
    6.66$\pm$   1.29 &     6.66$\pm$   1.29 & 
    2.40$\pm$   0.59 &     2.40$\pm$   0.59 & 
    19.5$\pm$    4.8 &     18.3$\pm$    4.5 \\ 
\foii\       7332.25 & 
--- & --- & 
--- & --- & 
    0.82$\pm$   0.25 &     0.82$\pm$   0.25 & 
--- & --- \\ 
 & & & & & & & & \\
F(\hb) $\times 10^{-16}$ erg s$^{-1}$ cm$^{-2}$ &
\multicolumn{2}{c|}{   3.63$\pm$   0.41} & 
\multicolumn{2}{c|}{   20.5$\pm$    2.1} & 
\multicolumn{2}{c|}{   36.1$\pm$    5.6} & 
\multicolumn{2}{c|}{   2.81$\pm$   0.30} \\ 
I(\hb) $\times 10^{-16}$ erg s$^{-1}$ cm$^{-2}$ &
\multicolumn{2}{c|}{   6.02$\pm$   0.81} & 
\multicolumn{2}{c|}{   20.5$\pm$    2.1} & 
\multicolumn{2}{c|}{   36.1$\pm$    5.9} & 
\multicolumn{2}{c|}{   3.27$\pm$   0.36} \\ 
E(B-V) &
\multicolumn{2}{c|}{ 0.15$\pm$ 0.02} & 
\multicolumn{2}{c|}{ 0.00$\pm$ 0.00} & 
\multicolumn{2}{c|}{ 0.00$\pm$ 0.01} & 
\multicolumn{2}{c|}{ 0.05$\pm$ 0.01} \\ 
\hline
\hline
\end{tabular}
\label{tab:fluxes_all}
\end{footnotesize}
\end{center}

\end{landscape}
\begin{landscape}

\begin{center}
\begin{footnotesize}
\begin{tabular}{|l|cc|cc|cc|cc|}
\hline
 & \multicolumn{2}{c|}{ KJ~5.0} & \multicolumn{2}{c|}{ KJ~7.0 }& \multicolumn{2}{c|}{ KJ~18.0} & \multicolumn{2}{c|}{ KJ~18.1} \\
\hline\hline
\foii\       3727.10 & 
    91.2$\pm$   13.8 &    110.0$\pm$   16.8 & 
     136$\pm$     19 &      148$\pm$     21 & 
     110$\pm$     16 &      128$\pm$     19 & 
     408$\pm$     70 &      528$\pm$     94 \\ 
\foii\       3729.86 & 
     105$\pm$     15 &      127$\pm$     19 & 
    78.0$\pm$   11.3 &     85.2$\pm$   12.4 & 
    86.7$\pm$   12.9 &    101.3$\pm$   15.2 & 
     179$\pm$     35 &      232$\pm$     47 \\ 
\hei\        3820.84 & 
    12.1$\pm$    2.6 &     14.3$\pm$    3.1 & 
--- & --- & 
    5.79$\pm$   1.71 &     6.66$\pm$   1.97 & 
     111$\pm$     21 &      140$\pm$     27 \\ 
\fneiii\       3870.16 & 
    8.43$\pm$   1.92 &     9.91$\pm$   2.26 & 
    26.7$\pm$    3.9 &     28.8$\pm$    4.2 & 
    23.3$\pm$    3.6 &     26.6$\pm$    4.1 & 
    38.5$\pm$   10.0 &     48.0$\pm$   12.6 \\ 
 \hei+H8       3889.75 & 
    7.60$\pm$   2.29 &     8.92$\pm$   2.69 & 
    14.5$\pm$    2.2 &     15.6$\pm$    2.4 & 
    15.9$\pm$    2.9 &     18.1$\pm$    3.3 & 
--- & --- \\ 
\heii\       3969.55 & 
    7.42$\pm$   1.94 &     8.58$\pm$   2.25 & 
    19.5$\pm$    2.9 &     20.9$\pm$    3.1 & 
    12.0$\pm$    2.2 &     13.5$\pm$    2.5 & 
--- & --- \\ 
\hei\        4027.49 & 
--- & --- & 
--- & --- & 
--- & --- & 
--- & --- \\ 
\fsii\       4069.75 & 
--- & --- & 
    2.90$\pm$   0.87 &     3.08$\pm$   0.93 & 
--- & --- & 
--- & --- \\ 
\hd\       4102.89 & 
    16.9$\pm$    2.9 &     19.0$\pm$    3.3 & 
    22.6$\pm$    3.3 &     24.0$\pm$    3.5 & 
    20.2$\pm$    3.2 &     22.3$\pm$    3.6 & 
--- & --- \\ 
\hg\       4341.68 & 
    37.3$\pm$    6.1 &     40.5$\pm$    6.7 & 
    41.1$\pm$    5.9 &     42.7$\pm$    6.1 & 
    39.3$\pm$    5.9 &     42.0$\pm$    6.4 & 
--- & --- \\ 
\foiii\        4364.44 & 
$<$   2.66  & $<$   2.88  & 
    4.86$\pm$   0.99 &     5.04$\pm$   1.02 & 
    6.45$\pm$   1.60 &     6.88$\pm$   1.71 & 
$<$   15.2  & $<$   16.9  \\ 
\hei\        4472.93 & 
    5.32$\pm$   1.68 &     5.65$\pm$   1.78 & 
    3.33$\pm$   0.76 &     3.43$\pm$   0.79 & 
--- & --- & 
--- & --- \\ 
\feiii\        4659.35 & 
--- & --- & 
--- & --- & 
--- & --- & 
--- & --- \\ 
\heii\        4687.02 & 
--- & --- & 
    2.08$\pm$   0.63 &     2.11$\pm$   0.64 & 
--- & --- & 
--- & --- \\ 
\ariv\        4741.49 & 
--- & --- & 
--- & --- & 
--- & --- & 
--- & --- \\ 
\hb\        4862.72 & 
   100.0$\pm$   14.4 &    100.0$\pm$   14.4 & 
     100$\pm$     14 &      100$\pm$     14 & 
     100$\pm$     14 &      100$\pm$     14 & 
     100$\pm$     18 &      100$\pm$     18 \\ 
\foiii\        4960.29 & 
    43.7$\pm$    6.5 &     43.0$\pm$    6.4 & 
     112$\pm$     16 &      112$\pm$     15 & 
    94.0$\pm$   13.5 &     92.8$\pm$   13.3 & 
    63.3$\pm$   19.5 &     62.0$\pm$   19.1 \\ 
\foiii\       5008.24 & 
     132$\pm$     19 &      129$\pm$     18 & 
     341$\pm$     48 &      338$\pm$     47 & 
     283$\pm$     40 &      279$\pm$     39 & 
     188$\pm$     34 &      183$\pm$     33 \\ 
\fnii\       5756.24 & 
--- & --- & 
--- & --- & 
--- & --- & 
--- & --- \\ 
\hei\        5877.59 & 
    10.5$\pm$    2.5 &      9.3$\pm$    2.2 & 
    9.33$\pm$   1.62 &     8.80$\pm$   1.53 & 
    11.0$\pm$    2.0 &      9.9$\pm$    1.8 & 
--- & --- \\ 
\foi\        6302.05 & 
--- & --- & 
    3.19$\pm$   0.68 &     2.95$\pm$   0.63 & 
    6.91$\pm$   1.47 &     6.02$\pm$   1.29 & 
--- & --- \\ 
\fsiii\       6313.80 & 
--- & --- & 
    1.72$\pm$   0.61 &     1.59$\pm$   0.57 & 
--- & --- & 
--- & --- \\ 
\foi\       6365.54 & 
--- & --- & 
    2.42$\pm$   0.60 &     2.24$\pm$   0.55 & 
--- & --- & 
--- & --- \\ 
\ha\       6564.61 & 
     309$\pm$     44 &      256$\pm$     37 & 
     292$\pm$     41 &      267$\pm$     37 & 
     308$\pm$     44 &      263$\pm$     37 & 
     371$\pm$     61 &      285$\pm$     48 \\ 
\fnii\        6585.20 & 
    7.01$\pm$   1.57 &     5.79$\pm$   1.31 & 
    13.0$\pm$    1.9 &     11.9$\pm$    1.8 & 
    8.53$\pm$   1.55 &     7.27$\pm$   1.33 & 
    44.1$\pm$   10.3 &     33.9$\pm$    8.0 \\ 
\hei\        6679.99 & 
--- & --- & 
    3.67$\pm$   0.68 &     3.35$\pm$   0.62 & 
    3.70$\pm$   0.99 &     3.14$\pm$   0.84 & 
    24.5$\pm$    7.6 &     18.6$\pm$    5.8 \\ 
\fsii\       6718.29 & 
    21.1$\pm$    3.3 &     17.2$\pm$    2.7 & 
    28.1$\pm$    4.0 &     25.6$\pm$    3.7 & 
    24.1$\pm$    3.6 &     20.4$\pm$    3.1 & 
    94.9$\pm$   17.0 &     71.9$\pm$   13.3 \\ 
\fsii\       6732.67 & 
    11.1$\pm$    1.9 &      9.0$\pm$    1.6 & 
    20.7$\pm$    3.0 &     18.8$\pm$    2.7 & 
    18.7$\pm$    2.8 &     15.8$\pm$    2.4 & 
    54.3$\pm$   11.0 &     41.1$\pm$    8.5 \\ 
\fneiii\        6870.16 & 
--- & --- & 
    1.35$\pm$   0.47 &     1.22$\pm$   0.43 & 
--- & --- & 
--- & --- \\ 
\fariii\       7137.80 & 
    5.74$\pm$   1.32 &     4.53$\pm$   1.05 & 
    6.52$\pm$   1.03 &     5.83$\pm$   0.92 & 
    6.58$\pm$   1.22 &     5.41$\pm$   1.01 & 
    22.1$\pm$    7.0 &     16.0$\pm$    5.1 \\ 
\foii\       7321.47 & 
--- & --- & 
--- & --- & 
--- & --- & 
--- & --- \\ 
\foii\       7332.25 & 
    6.55$\pm$   1.84 &     5.09$\pm$   1.44 & 
--- & --- & 
    4.10$\pm$   1.12 &     3.33$\pm$   0.91 & 
    27.6$\pm$    8.8 &     19.5$\pm$    6.3 \\ 
 & & & & & & & & \\
F(\hb) $\times 10^{-16}$ erg s$^{-1}$ cm$^{-2}$ &
\multicolumn{2}{c|}{   7.06$\pm$   0.72} & 
\multicolumn{2}{c|}{   19.2$\pm$    1.9} & 
\multicolumn{2}{c|}{   6.74$\pm$   0.68} & 
\multicolumn{2}{c|}{   1.43$\pm$   0.19} \\ 
I(\hb) $\times 10^{-16}$ erg s$^{-1}$ cm$^{-2}$ &
\multicolumn{2}{c|}{   13.0$\pm$    1.7} & 
\multicolumn{2}{c|}{   25.5$\pm$    2.7} & 
\multicolumn{2}{c|}{   11.1$\pm$    1.3} & 
\multicolumn{2}{c|}{   3.27$\pm$   0.61} \\ 
E(B-V) &
\multicolumn{2}{c|}{ 0.18$\pm$ 0.02} & 
\multicolumn{2}{c|}{ 0.09$\pm$ 0.01} & 
\multicolumn{2}{c|}{ 0.15$\pm$ 0.02} & 
\multicolumn{2}{c|}{ 0.25$\pm$ 0.04} \\ 
\hline
\hline
\end{tabular}
\label{tab:fluxes_all}
\end{footnotesize}
\end{center}

\end{landscape}
\begin{landscape}
\begin{center}
\begin{footnotesize}
\begin{tabular}{|l|cc|cc|cc|cc|}
\hline
 & \multicolumn{2}{c|}{ KJ~22.0} & \multicolumn{2}{c|}{ KJ~22.1 }& \multicolumn{2}{c|}{ KJ~29.0} & \multicolumn{2}{c|}{ KJ~29.2} \\
\hline\hline
\foii\       3727.10 & 
    66.4$\pm$    9.5 &     76.1$\pm$   11.0 & 
    64.2$\pm$    9.3 &     73.6$\pm$   10.7 & 
    87.7$\pm$   13.6 &     95.9$\pm$   15.0 & 
    87.3$\pm$   13.3 &    102.4$\pm$   15.7 \\ 
\foii\       3729.86 & 
    71.6$\pm$   10.3 &     82.0$\pm$   11.8 & 
    80.2$\pm$   11.5 &     92.0$\pm$   13.3 & 
    97.6$\pm$   15.0 &    106.7$\pm$   16.5 & 
    79.4$\pm$   12.2 &     93.1$\pm$   14.4 \\ 
\hei\        3820.84 & 
--- & --- & 
--- & --- & 
--- & --- & 
    7.19$\pm$   2.14 &     8.31$\pm$   2.47 \\ 
\fneiii\       3870.16 & 
    32.3$\pm$    4.6 &     36.3$\pm$    5.2 & 
    31.1$\pm$    4.5 &     35.0$\pm$    5.1 & 
    10.4$\pm$    2.5 &     11.3$\pm$    2.7 & 
     110$\pm$     16 &      126$\pm$     18 \\ 
 \hei+H8       3889.75 & 
    13.9$\pm$    2.1 &     15.6$\pm$    2.3 & 
    13.4$\pm$    2.0 &     15.0$\pm$    2.3 & 
    12.4$\pm$    2.8 &     13.4$\pm$    3.0 & 
--- & --- \\ 
\heii\       3969.55 & 
    21.2$\pm$    3.1 &     23.6$\pm$    3.4 & 
    20.5$\pm$    3.0 &     22.8$\pm$    3.4 & 
--- & --- & 
    11.0$\pm$    2.5 &     12.4$\pm$    2.8 \\ 
\hei\        4027.49 & 
    2.15$\pm$   0.62 &     2.38$\pm$   0.68 & 
--- & --- & 
--- & --- & 
--- & --- \\ 
\fsii\       4069.75 & 
--- & --- & 
--- & --- & 
--- & --- & 
--- & --- \\ 
\hd\       4102.89 & 
    21.7$\pm$    3.1 &     23.7$\pm$    3.4 & 
    22.3$\pm$    3.3 &     24.4$\pm$    3.6 & 
    12.0$\pm$    2.6 &     12.7$\pm$    2.8 & 
    19.5$\pm$    3.3 &     21.7$\pm$    3.7 \\ 
\hg\       4341.68 & 
    43.0$\pm$    6.1 &     45.6$\pm$    6.5 & 
    41.3$\pm$    5.9 &     43.8$\pm$    6.3 & 
    44.2$\pm$    6.8 &     45.9$\pm$    7.1 & 
    41.3$\pm$    6.3 &     44.3$\pm$    6.8 \\ 
\foiii\        4364.44 & 
    7.19$\pm$   1.16 &     7.61$\pm$   1.22 & 
    7.18$\pm$   1.25 &     7.60$\pm$   1.32 & 
$<$   3.44  & $<$   3.57  & 
$<$   2.76  & $<$   2.95  \\ 
\hei\        4472.93 & 
    3.75$\pm$   0.70 &     3.92$\pm$   0.73 & 
    4.50$\pm$   0.89 &     4.70$\pm$   0.93 & 
--- & --- & 
--- & --- \\ 
\feiii\        4659.35 & 
--- & --- & 
--- & --- & 
--- & --- & 
--- & --- \\ 
\heii\        4687.02 & 
    2.11$\pm$   0.55 &     2.16$\pm$   0.56 & 
--- & --- & 
    6.36$\pm$   1.95 &     6.45$\pm$   1.97 & 
--- & --- \\ 
\ariv\        4741.49 & 
--- & --- & 
--- & --- & 
--- & --- & 
--- & --- \\ 
\hb\        4862.72 & 
     100$\pm$     14 &      100$\pm$     14 & 
     100$\pm$     14 &      100$\pm$     14 & 
     100$\pm$     14 &      100$\pm$     14 & 
     100$\pm$     14 &      100$\pm$     14 \\ 
\foiii\        4960.29 & 
     171$\pm$     24 &      169$\pm$     24 & 
     153$\pm$     21 &      151$\pm$     21 & 
    51.2$\pm$    7.7 &     50.9$\pm$    7.7 & 
    63.2$\pm$    9.3 &     62.4$\pm$    9.2 \\ 
\foiii\       5008.24 & 
     512$\pm$     72 &      504$\pm$     71 & 
     458$\pm$     64 &      451$\pm$     64 & 
     151$\pm$     22 &      149$\pm$     21 & 
     182$\pm$     26 &      179$\pm$     25 \\ 
\fnii\       5756.24 & 
--- & --- & 
--- & --- & 
--- & --- & 
--- & --- \\ 
\hei\        5877.59 & 
    11.1$\pm$    1.7 &     10.1$\pm$    1.5 & 
    12.6$\pm$    2.0 &     11.5$\pm$    1.8 & 
    8.81$\pm$   2.49 &     8.30$\pm$   2.35 & 
    12.7$\pm$    2.6 &     11.4$\pm$    2.4 \\ 
\foi\        6302.05 & 
    3.60$\pm$   0.70 &     3.19$\pm$   0.62 & 
    2.09$\pm$   0.68 &     1.85$\pm$   0.61 & 
--- & --- & 
--- & --- \\ 
\fsiii\       6313.80 & 
    2.00$\pm$   0.57 &     1.77$\pm$   0.50 & 
--- & --- & 
--- & --- & 
--- & --- \\ 
\foi\       6365.54 & 
    1.49$\pm$   0.49 &     1.31$\pm$   0.43 & 
--- & --- & 
--- & --- & 
    4.27$\pm$   1.48 &     3.69$\pm$   1.28 \\ 
\ha\       6564.61 & 
     322$\pm$     45 &      281$\pm$     40 & 
     313$\pm$     44 &      272$\pm$     38 & 
     309$\pm$     44 &      282$\pm$     41 & 
     322$\pm$     46 &      274$\pm$     39 \\ 
\fnii\        6585.20 & 
    8.65$\pm$   1.35 &     7.52$\pm$   1.18 & 
    6.84$\pm$   1.20 &     5.94$\pm$   1.05 & 
    6.42$\pm$   1.94 &     5.86$\pm$   1.77 & 
$<$   2.58  & $<$   2.19  \\ 
\hei\        6679.99 & 
    4.35$\pm$   0.77 &     3.76$\pm$   0.67 & 
    3.92$\pm$   0.82 &     3.38$\pm$   0.71 & 
    11.8$\pm$    3.3 &     10.7$\pm$    3.0 & 
    10.2$\pm$    2.9 &      8.6$\pm$    2.5 \\ 
\fsii\       6718.29 & 
    21.6$\pm$    3.1 &     18.7$\pm$    2.7 & 
    19.0$\pm$    2.8 &     16.4$\pm$    2.4 & 
    13.6$\pm$    2.8 &     12.3$\pm$    2.5 & 
    11.1$\pm$    2.3 &      9.4$\pm$    2.0 \\ 
\fsii\       6732.67 & 
    13.9$\pm$    2.0 &     12.0$\pm$    1.8 & 
    13.6$\pm$    2.0 &     11.7$\pm$    1.8 & 
    11.4$\pm$    2.5 &     10.3$\pm$    2.3 & 
    14.1$\pm$    2.6 &     11.8$\pm$    2.2 \\ 
\fneiii\        6870.16 & 
--- & --- & 
--- & --- & 
--- & --- & 
--- & --- \\ 
\fariii\       7137.80 & 
    7.11$\pm$   1.12 &     5.99$\pm$   0.95 & 
    6.76$\pm$   1.13 &     5.68$\pm$   0.95 & 
--- & --- & 
--- & --- \\ 
\foii\       7321.47 & 
    2.62$\pm$   0.67 &     2.18$\pm$   0.56 & 
--- & --- & 
    5.38$\pm$   1.95 &     4.77$\pm$   1.73 & 
--- & --- \\ 
\foii\       7332.25 & 
    3.52$\pm$   0.67 &     2.93$\pm$   0.56 & 
    2.79$\pm$   0.70 &     2.32$\pm$   0.58 & 
--- & --- & 
--- & --- \\ 
 & & & & & & & & \\
F(\hb) $\times 10^{-16}$ erg s$^{-1}$ cm$^{-2}$ &
\multicolumn{2}{c|}{   24.0$\pm$    2.4} & 
\multicolumn{2}{c|}{   15.4$\pm$    1.5} & 
\multicolumn{2}{c|}{   4.44$\pm$   0.46} & 
\multicolumn{2}{c|}{   5.68$\pm$   0.59} \\ 
I(\hb) $\times 10^{-16}$ erg s$^{-1}$ cm$^{-2}$ &
\multicolumn{2}{c|}{   37.2$\pm$    4.2} & 
\multicolumn{2}{c|}{   24.0$\pm$    2.7} & 
\multicolumn{2}{c|}{   5.94$\pm$   0.65} & 
\multicolumn{2}{c|}{   9.51$\pm$   1.15} \\ 
E(B-V) &
\multicolumn{2}{c|}{ 0.13$\pm$ 0.01} & 
\multicolumn{2}{c|}{ 0.13$\pm$ 0.02} & 
\multicolumn{2}{c|}{ 0.09$\pm$ 0.01} & 
\multicolumn{2}{c|}{ 0.16$\pm$ 0.02} \\ 
\hline
\hline
\end{tabular}
\label{tab:fluxes_all}
\end{footnotesize}
\end{center}

\end{landscape}
\begin{landscape}
\begin{center}
\begin{footnotesize}
\begin{tabular}{|l|cc|cc|cc|cc|}
\hline
 & \multicolumn{2}{c|}{ KJ~37.0} & \multicolumn{2}{c|}{ KJ~53.0 }& \multicolumn{2}{c|}{ KJ~78.0}  \\
\hline\hline
\foii\       3727.10 & 
    97.0$\pm$   14.0 &    130.0$\pm$   19.2 & 
    41.8$\pm$    6.0 &     42.0$\pm$    6.0 & 
    44.7$\pm$    6.6 &     45.1$\pm$    6.7 \\ 
\foii\       3729.86 & 
    75.6$\pm$   11.0 &    101.3$\pm$   15.1 & 
    42.8$\pm$    6.1 &     43.0$\pm$    6.1 & 
    46.9$\pm$    6.9 &     47.3$\pm$    7.0 \\ 
\hei\        3820.84 & 
--- & --- & 
--- & --- & 
--- & --- \\ 
\fneiii\       3870.16 & 
    26.5$\pm$    3.9 &     34.0$\pm$    5.1 & 
    26.0$\pm$    3.7 &     26.1$\pm$    3.7 & 
    22.4$\pm$    3.3 &     22.6$\pm$    3.3 \\ 
 \hei+H8       3889.75 & 
    12.0$\pm$    2.1 &     15.4$\pm$    2.7 & 
    17.9$\pm$    2.6 &     18.0$\pm$    2.6 & 
    16.8$\pm$    2.5 &     16.9$\pm$    2.5 \\ 
\heii\       3969.55 & 
    14.4$\pm$    2.4 &     18.1$\pm$    3.0 & 
    20.0$\pm$    2.8 &     20.1$\pm$    2.9 & 
    20.7$\pm$    3.0 &     20.9$\pm$    3.1 \\ 
\hei\        4027.49 & 
--- & --- & 
    1.24$\pm$   0.27 &     1.25$\pm$   0.27 & 
    2.48$\pm$   0.70 &     2.49$\pm$   0.71 \\ 
\fsii\       4069.75 & 
--- & --- & 
    0.80$\pm$   0.23 &     0.80$\pm$   0.23 & 
    2.23$\pm$   0.73 &     2.24$\pm$   0.73 \\ 
\hd\       4102.89 & 
    13.8$\pm$    2.3 &     16.7$\pm$    2.7 & 
    24.1$\pm$    3.4 &     24.1$\pm$    3.4 & 
    23.3$\pm$    3.4 &     23.4$\pm$    3.4 \\ 
\hg\       4341.68 & 
    38.8$\pm$    5.6 &     44.1$\pm$    6.4 & 
    45.1$\pm$    6.4 &     45.1$\pm$    6.4 & 
    41.8$\pm$    6.0 &     41.9$\pm$    6.0 \\ 
\foiii\        4364.44 & 
    7.63$\pm$   1.57 &     8.62$\pm$   1.77 & 
    8.93$\pm$   1.28 &     8.95$\pm$   1.29 & 
    7.46$\pm$   1.27 &     7.49$\pm$   1.27 \\ 
\hei\        4472.93 & 
--- & --- & 
    3.18$\pm$   0.48 &     3.19$\pm$   0.48 & 
    3.15$\pm$   0.69 &     3.16$\pm$   0.69 \\ 
\feiii\        4659.35 & 
--- & --- & 
--- & --- & 
--- & --- \\ 
\heii\        4687.02 & 
--- & --- & 
--- & --- & 
--- & --- \\ 
\ariv\        4741.49 & 
--- & --- & 
--- & --- & 
--- & --- \\ 
\hb\        4862.72 & 
     100$\pm$     14 &      100$\pm$     14 & 
     100$\pm$     14 &       99$\pm$     14 & 
     100$\pm$     14 &       99$\pm$     14 \\ 
\foiii\        4960.29 & 
     150$\pm$     21 &      146$\pm$     20 & 
     113$\pm$     15 &      113$\pm$     15 & 
     117$\pm$     16 &      117$\pm$     16 \\ 
\foiii\       5008.24 & 
     419$\pm$     59 &      406$\pm$     57 & 
     338$\pm$     47 &      338$\pm$     47 & 
     357$\pm$     50 &      357$\pm$     50 \\ 
\fnii\       5756.24 & 
--- & --- & 
--- & --- & 
--- & --- \\ 
\hei\        5877.59 & 
    10.9$\pm$    2.2 &      8.9$\pm$    1.8 & 
   10.00$\pm$   1.43 &     9.98$\pm$   1.43 & 
    9.75$\pm$   1.60 &     9.69$\pm$   1.59 \\ 
\foi\        6302.05 & 
    6.69$\pm$   1.42 &     5.16$\pm$   1.10 & 
    0.85$\pm$   0.22 &     0.84$\pm$   0.22 & 
--- & --- \\ 
\fsiii\       6313.80 & 
    5.69$\pm$   1.34 &     4.38$\pm$   1.04 & 
    1.30$\pm$   0.25 &     1.30$\pm$   0.25 & 
--- & --- \\ 
\foi\       6365.54 & 
--- & --- & 
--- & --- & 
--- & --- \\ 
\ha\       6564.61 & 
     368$\pm$     52 &      273$\pm$     40 & 
     280$\pm$     39 &      279$\pm$     39 & 
     265$\pm$     37 &      263$\pm$     37 \\ 
\fnii\        6585.20 & 
    12.5$\pm$    2.1 &      9.2$\pm$    1.6 & 
    2.83$\pm$   0.44 &     2.82$\pm$   0.44 & 
    4.68$\pm$   0.82 &     4.64$\pm$   0.81 \\ 
\hei\        6679.99 & 
--- & --- & 
    3.02$\pm$   0.46 &     3.01$\pm$   0.45 & 
    2.54$\pm$   0.53 &     2.52$\pm$   0.52 \\ 
\fsii\       6718.29 & 
    34.0$\pm$    5.0 &     24.8$\pm$    3.7 & 
    6.83$\pm$   0.98 &     6.80$\pm$   0.98 & 
    6.31$\pm$   1.00 &     6.25$\pm$   0.99 \\ 
\fsii\       6732.67 & 
    22.8$\pm$    3.5 &     16.6$\pm$    2.6 & 
    5.04$\pm$   0.73 &     5.02$\pm$   0.73 & 
    5.33$\pm$   0.87 &     5.28$\pm$   0.86 \\ 
\fneiii\        6870.16 & 
--- & --- & 
--- & --- & 
--- & --- \\ 
\fariii\       7137.80 & 
    10.8$\pm$    2.0 &      7.5$\pm$    1.4 & 
    4.03$\pm$   0.59 &     4.02$\pm$   0.59 & 
    5.00$\pm$   0.83 &     4.95$\pm$   0.82 \\ 
\foii\       7321.47 & 
--- & --- & 
    1.21$\pm$   0.26 &     1.20$\pm$   0.25 & 
--- & --- \\ 
\foii\       7332.25 & 
    3.89$\pm$   1.18 &     2.63$\pm$   0.81 & 
    1.60$\pm$   0.34 &     1.59$\pm$   0.34 & 
    2.94$\pm$   0.85 &     2.91$\pm$   0.84 \\ 
 & & & & & & & & \\
F(\hb) $\times 10^{-16}$ erg s$^{-1}$ cm$^{-2}$ &
\multicolumn{2}{c|}{   9.11$\pm$   0.92} & 
\multicolumn{2}{c|}{   62.3$\pm$    6.2} & 
\multicolumn{2}{c|}{   26.7$\pm$    2.7} \\ 
I(\hb) $\times 10^{-16}$ erg s$^{-1}$ cm$^{-2}$ &
\multicolumn{2}{c|}{   23.4$\pm$    3.5} & 
\multicolumn{2}{c|}{   63.1$\pm$    6.3} & 
\multicolumn{2}{c|}{   27.4$\pm$    2.8} \\ 
E(B-V) &
\multicolumn{2}{c|}{ 0.29$\pm$ 0.03} & 
\multicolumn{2}{c|}{ 0.00$\pm$ 0.00} & 
\multicolumn{2}{c|}{ 0.01$\pm$ 0.00} \\ 
\hline
\hline
\end{tabular}
\label{tab:fluxes_all}
\end{footnotesize}
\end{center}

\end{landscape}
\bsp

\label{lastpage}

\end{document}